\documentclass[aps,
               pra,
               twocolumn,
               nofootinbib,
               floatfix,
               showpacs,
	      superscriptaddress
              ]{revtex4}

\usepackage{float}
\usepackage{amssymb}
\usepackage{amsmath}
\usepackage{graphics}
\usepackage{pstricks}
\usepackage{bm}
\usepackage{graphicx}
\usepackage{color}

\renewcommand{\i}{\mathrm{i}}
\renewcommand{\v}{\mathrm{V}}
\renewcommand{\a}{\mathrm{A}}

\renewcommand{\vector}[1]{\mathbf{#1}}

\newcommand{\ddf}[1]{\mathrm{d}^{d}\! #1 \,}
\renewcommand{\vec}{\mathbf}
\newcommand{\cred}[1]{{\color{red}{#1}}}

\newcommand{\weg}[1]{{\cred{#1}}}

\renewcommand{\weg}[1]{}

\newcommand{\del}{\partial}

\newcommand{\eq}[1]{(\ref{eq:#1})}
\newcommand{\Eq}[1]{Eq.~(\ref{eq:#1})}

\newcommand{\Fig}[1]{Fig.~\ref{fig:#1}}
\newcommand{\fig}[1]{\ref{fig:#1}}
\newcommand{\Sectionref}[1]{Section~\ref{sec:#1}}
\newcommand{\Sect}[1]{Sect.~\ref{sec:#1}}
\newcommand{\Subsect}[1]{Sect.~\ref{subsec:#1}}

\newcommand{\subsect}[1]{\ref{subsec:#1}}
\newcommand{\Appendix}[1]{Appendix \ref{app:#1}}
\newcommand{\App}[1]{App.~\ref{app:#1}}

\begin{document}

\title{Nonthermal fixed points, vortex statistics, and superfluid turbulence\\ in an ultracold Bose gas}

\author{Boris~Nowak}
%\email{b.nowak@thphys.uni-heidelberg.de}
\affiliation{Institut f\"ur Theoretische Physik,
             Ruprecht-Karls-Universit\"at Heidelberg,
             Philosophenweg~16,
             69120~Heidelberg, Germany}
\affiliation{ExtreMe Matter Institute EMMI,
             GSI Helmholtzzentrum f\"ur Schwerionenforschung GmbH, 
             Planckstra\ss e~1, 
             64291~Darmstadt, Germany} 
\affiliation{Kavli Institute for Theoretical Physics,
	      University of California,
	      Santa Barbara, CA 93106, USA}
\author{Jan~Schole}
%\email{j.schole@thphys.uni-heidelberg.de}
\affiliation{Institut f\"ur Theoretische Physik,
             Ruprecht-Karls-Universit\"at Heidelberg,
             Philosophenweg~16,
             69120~Heidelberg, Germany}
\affiliation{ExtreMe Matter Institute EMMI,
             GSI Helmholtzzentrum f\"ur Schwerionenforschung GmbH, 
             Planckstra\ss e~1, 
             64291~Darmstadt, Germany} 
\author{D\'enes~Sexty}
%\email{d.sexty@thphys.uni-heidelberg.de}
\author{Thomas~Gasenzer}
\email{t.gasenzer@uni-heidelberg.de}
\affiliation{Institut f\"ur Theoretische Physik,
             Ruprecht-Karls-Universit\"at Heidelberg,
             Philosophenweg~16,
             69120~Heidelberg, Germany}
\affiliation{ExtreMe Matter Institute EMMI,
             GSI Helmholtzzentrum f\"ur Schwerionenforschung GmbH, 
             Planckstra\ss e~1, 
             64291~Darmstadt, Germany} 
\affiliation{Kavli Institute for Theoretical Physics,
	      University of California,
	      Santa Barbara, CA 93106, USA}

\date{\today}

\begin{abstract}

Nonthermal fixed points of the dynamics of a dilute degenerate Bose gas far from thermal equilibrium are analysed in two and three spatial dimensions. 
Universal power-law distributions, previously found within a nonperturbative quantum-field theoretical approach and recently shown to be related to vortical dynamics and superfluid turbulence [Phys.~Rev.~B \textbf{84}, 020506(R) (2011)], are studied in detail. 
The results imply an interpretation of the scaling behavior in terms of independent vortex excitations of the superfluid and show that the statistics of topological excitations can be described in the framework of wave turbulence. 
The particular scaling exponents observed in the single-particle momentum distributions are found to be consistent with irreversibility as well as conservation laws obeyed by the wave interactions.
Moreover, long-wavelength acoustic excitations of the vortex-bearing condensate, driven by vortex annihilations, are found to follow a nonthermal power law.
Considering vortex correlations in a statistical model, the long-time departure from the nonthermal fixed point is related to vortex-antivortex pairing.
The studied nonthermal fixed points are accessible in cold-gas experiments.
The results shed light on fundamental aspects of superfluid turbulence and have strong potential implications for related phenomena, e.g., in early-universe inflation or quark-gluon plasma dynamics.
\end{abstract}

% insert suggested PACS numbers in braces on next line

\maketitle

%== Introduction ================================================================
\section{Introduction}
\label{sec:intro}

Turbulence is a generic phenomenon observed in the relaxation dynamics of many-body systems far from thermal equilibrium \cite{Frisch1995a}.
It comprises a quasi-stationary flow of energy within certain inertial regimes in momentum space \cite{Richardson1920a}.
Correlation functions like the energy spectrum and higher momenta of the velocity distribution exhibit universality and scaling \cite{Kolmogorov1941a, Obukhov1941a}. 
This, and quasistationarity are the key characteristics rendering turbulence a nonthermal fixed point of the system's dynamics.

In quantum many-body systems, from the formation of Bose-Einstein condensates in ultracold gases to quark-gluon plasmas produced in heavy-ion
collisions and reheating after early-universe inflation, nonequilibrium dynamics governs many interesting phenomena.
In this context, turbulence during thermalization is being studied with increasing effort \cite{Micha:2002ey, Arnold:2005ef, Arnold:2005qs, Mueller:2006up, Berges:2008wm,Berges:2008sr, Berges:2008mr, Scheppach:2009wu, Berges:2010ez, Nowak:2010tm, Carrington:2010sz, Fukushima2011a, Gasenzer:2011by}. 
A nonthermal fixed point of the evolution of a many-body system has the potential to strongly affect the equilibration process by forcing the evolution to critically slow down before the system can thermalize.
New scaling laws were found by analyzing non-perturbative quantum field dynamic equations  \cite{Berges:2008wm,Berges:2008sr,Berges:2008mr,Berges:2010ez}.
Analogous predictions for a dilute ultracold Bose gas were given in \cite{Scheppach:2009wu}, proposing strong matter-wave turbulence in the regime of long-range excitations.

Superfluid turbulence, also referred to as quantum turbulence (QT) has been the subject of extensive studies in the context of helium \cite{Halperin2008a, Donnelly1991a} and dilute Bose gases~\cite{Horng2007a,Horng2009a,Foster2010a,Yukalov2010a}. 
In contrast to eddies in classical fluids, vorticity in a superfluid is quantized \cite{Onsager1949a,Feynman1955a}, and the creation and annihilation processes of quantized vortices are distinctly different \cite{Halperin2008a,Donnelly1991a}. 
The observation of a Kolmogorov 5/3-law \cite{Kolmogorov1941a,Obukhov1941a} in experiments with superfluid helium~\cite{Maurer1998, Stalp1999, Vinen2002a} received much attention 
\cite{Nore1997a,Nore1997b,Araki2002a,Kobayashi2005a,Kobayashi2005b,Tsubota2008a,krstulovic2011}. 
In particular, the role of the normal-fluid as compared to the superfluid component in the turbulent flow is under debate ~\cite{Halperin2008a, Donnelly1991a}.

Superfluid turbulence plays an important role in the context of the kinetics of condensation and the development of long-range order in a dilute Bose gas. 
This, as well as turbulence in its acoustic excitations has been discussed in Refs.~\cite{Levich1978a,Kagan1992a,Kagan1994a,Kagan1997c,Berloff2002a,Svistunov2001a,Kozik2009a}.
A possible observation of QT in ultracold atomic gases presently poses an exciting task for experiments ~\cite{Weiler2008a,Henn2009a,seman2011}. 
Here we emphasize that the experimental study of superfluid turbulence and, more generally, of nonthermal fixed points in ultracold Bose gases has strong potential implications for many other areas of physics.
Besides vortical excitations this also includes other (quasi-)topological excitations such as solitary waves in one-dimensional gases.

A satisfactory ab-initio mathematical description of both quantum and classical turbulence is inherently difficult due to the strong correlations building up within the system. 
Analytical results are known, however, in regimes where kinetic theory applies: 
In a dilute, degenerate Bose gas the normal-fluid component can vary at the expense or gain of the superfluid part. 
As a consequence, the gas is compressible and so-called weak wave-turbulence phenomena can occur for which scaling laws can be derived by analyzing kinetic equations \cite{Zakharov1992a, Nazarenko2011a,Newell2011a}. 

Generically, however, the description in terms of wave kinetic equations such as the Quantum Boltzmann equation breaks down in the infrared (IR) regime of long wavelengths. For a Bose gas in this regime, single-particle occupation numbers grow large and the description in terms of, for example, elastic two-to-two collisions becomes unreliable.
In the infrared limit, so-called strong wave turbulence is expected to occur.
Recent developments presented in Refs.~\cite{Berges:2008wm,Berges:2008sr,Berges:2008mr,Scheppach:2009wu,Berges:2010ez,Carrington:2010sz} allow one to set up a unifying description of scaling, both in the ultraviolet (UV) Quantum Boltzmann kinetic regime and in the infrared limit. 
In the IR regime, new scaling laws were found for a relativistic scalar field by analyzing non-perturbative Kadanoff-Baym dynamic equations with respect to nonthermal stationary solutions ~\cite{Berges:2008wm}. 

In this article, we present the details of our studies of the relaxation of two and three dimensional dilute Bose gases through stages of superfluid turbulence and the approach of a nonthermal fixed point, by means of simulations in the classical-wave limit of the underlying quantum field theory. 
In \Sectionref{dyncritpoints}, we summarize the quantum-field theoretical predictions of Refs.~\cite{Zakharov1992a,Scheppach:2009wu}, in particular for the scaling exponents of the single-particle momentum distribution and compare these with the numerically determined scaling.
While we find excellent agreement, our results provide an interpretation of the nonthermal fixed points proposed in Refs.~\cite{Berges:2008wm,Scheppach:2009wu} for the case of an ultracold Bose gas: 
The appearance of nonperturbative infrared scaling reflects the presence of statistically independent vortices, as previously reported in brief in Ref.~\cite{Nowak:2010tm}, see also~Figs.~\fig{2DPhase512}-\fig{ModeOccupation3g8}. 
This phenomenon appears to be distinctly different from the weak wave turbulence we observe in the UV part of the spectrum.

In \Sectionref{RandomVortexModel}, we describe a model of independent as well as pair-correlated vortical excitations \cite{Onsager1949a} which allows the more refined interpretation of the scaling behavior during the different stages of the evolution presented in \Sectionref{Results}.
We show that the stationary scaling is maintained by the presence of energy (UV) and particle (IR) fluxes, further supporting the analytic theory. 
In \Sectionref{VorticesAnd}, we study the power-law distributions of the compressible and incompressible contributions to the flow pattern.
This adds to the clear understanding of the bimodal scaling laws in the overall momentum spectrum.
IR power-law spectra of subdominant compressible excitations suggest the presence of acoustic turbulence \cite{Zakharov1992a} on the top of the vorticity-bearing quasicondensate. 
In comparison with recent experiments and analytical predictions, the velocity field probability distribution as well as the velocity statistics of individual vortices are studied. 
This observable is of great interest, since it has recently been used to experimentally verify a distinction between classical and quantum turbulence \cite{Paoletti2008a}.
We finally show that the complete decay of the turbulent scaling is anticipated by the appearance of weaker IR power laws reflecting vortex-antivortex pairing correlations. 
\\

%==Superfluid turbulence as a nonthermal fixed point ================================================================
%======================================================================
\section{Superfluid turbulence as a nonthermal fixed point}
\label{sec:dyncritpoints}

We begin with a brief summary of the analytical and numerical results on dynamical fixed points and matter-wave turbulence reported in Refs.~\cite{Scheppach:2009wu, Nowak:2010tm} before we discuss these results in terms of a statistical model of vortex excitations.
Motivated by the original work presented in Ref.~\cite{Berges:2008wm} nonthermal fixed points of Kadanoff-Baym dynamic equations for time-dependent Green functions were analysed, for the case of an ultracold Bose gas, in Ref.~\cite{Scheppach:2009wu}. At these fixed points of the dynamical evolution of the system, the single-particle momentum distribution $n(\mathbf{k},t)$ becomes (quasi-)stationary, $\dot{n}(\vector{k},t)=0$. It furthermore exhibits a characteristic universal power-law behavior, i.e., it scales according to
\begin{equation}
n(s\vector{k}) = s^{-\zeta}n(\vector{k})\,,
 \label{eq:scale}
\end{equation}
in a certain regime of momenta $\mathbf{k}$. Here $s$ is some positive, real number and $\zeta$ a universal exponent which was determined from the dynamic equations for the field correlation functions. Different exponents resulted in different momentum regimes.
Exponents valid in the ultraviolet regime of large $|\mathbf{k}|$ were found to correspond to well-known fixed points of weak wave turbulence \cite{Zakharov1992a} in the respective systems. 
In addition to these, new, larger exponents were predicted in the infrared regime of small $|\mathbf{k}|$ on the basis of a non-perturbative analysis of the Kadanoff-Baym dynamic equations derived from the two-particle-irreducible (2PI) effective action  \cite{Berges:2008wm,Berges:2008sr, Berges:2008mr, Scheppach:2009wu, Berges:2010ez}.

Simulations of the classical field equations for a relativistic $O(N)$-symmetric scalar model, for $N=4$, confirmed the existence of the analytically derived scalings in the infrared regime~\cite{Berges:2008wm}. 
As reported in Ref.~\cite{Nowak:2010tm}, corresponding simulations were performed for an ultracold, degenerate Bose gas in two and three spatial dimensions which demonstrated that for these systems the infrared scalings predicted in \cite{Scheppach:2009wu} reflect and are caused by the presence of quantized vortical excitations of the superfluid. 
In this way, the nonthermal fixed points can be related in a clear manner to topological excitations of the interacting coherent matter-wave field which shows that both, weak wave turbulence and macroscopic, topological excitations of the field can be described within a unified field-theoretic approach.
Differently expressed, both, weak turbulent flow and non-linear solitary bulk excitations are described in a unified manner as representing a nonthermal critical fixed point of the system.
In turn, the approach also implies that superfluid turbulence can be studied in a new way, in the frame of a universal quantum field theoretical approach.
We remark that the relation between the new infrared exponents and topological excitations of the background (i.e., condensate) field gains further support by the topological pattern formation described in Ref.~\cite{Gasenzer:2011by} for a relativistic $O(2)$-symmetric scalar model, which serves as a particular model for the reheating period after early-universe inflation.

We will discuss in detail vortical excitations in a degenerate Bose gas and their relation to the above nonthermal fixed points.
Before this we summarize briefly the relevant results of Refs. \cite{Scheppach:2009wu, Nowak:2010tm}. To be specific, throughout this article we consider an ultracold Bose gas of atoms in $d=2$, $3$ space dimensions interacting through $s$-wave collisions which is described by the Hamiltonian
\begin{equation}
 H =   \int  \mathrm{d}^dx \, \left[ -\Phi^\dag\frac{\nabla^2}{2m}\Phi  +  \frac{g}{2} \, \Phi^\dag\Phi^\dag \Phi \Phi \right] \,,
\end{equation}
where the time and space dependent fields $\Phi \equiv \Phi(\mathbf{x},t) $ satisfy Bose commutation relations, and where the coupling $g=4\pi a/m$ in three dimensions is defined in terms of the $s$-wave scattering length $a$. 
Here and in the following we set $\hbar=1$, see \Appendix{classfields} for more details.

%======================================================================
\subsection{Weak wave turbulence}
\label{subsec:wwt}
Suppose the generic case that for sufficiently large momenta $|\mathbf{k}|=k$ occupation numbers $n(\mathbf{k},t) = \langle\Phi^{\dagger}(\mathbf{k},t) \Phi(\mathbf{k},t) \rangle$ are small enough such that for a given coupling $g$ the Quantum Boltzmann Equation (QBE) 
\begin{align}
  \partial_{t}n_{\vec k}
  &= I(\vec k,t),
  \label{eq:QKinEq}
  \\
   I(\vec k,t)
  &= \int\ddf{p}\ddf{q}\ddf{r}|T_{\vec k\vec p\vec q\vec r}|^{2}\delta(\vec k+\vec p -\vec q - \vec r)
  \nonumber\\
  &\qquad\quad\times\
  \delta(\omega_{\vec k}+\omega_{\vec p}-\omega_{\vec q}-\omega_{\vec r})
  \nonumber\\
  &\qquad\quad\times\
  [(n_{\vec k}+1)(n_{\vec p}+1)n_{\vec q}n_{\vec r}
  \nonumber\\
  &\qquad\qquad -\
  n_{\vec k}n_{\vec p}(n_{\vec q}+1)(n_{\vec r}+1)],
  \label{eq:KinScattInt}
\end{align}
describes the evolution of $n_{\mathbf{k}}\equiv n(\mathbf{k},t)$ under the effects of collisions. 
In our case, the transition matrix element squared $|T_{\vec p\vec k\vec q\vec r}|^{2}$ is a numerical constant proportional to $g^{2}$ and thus independent of momenta.
Zeroes of the scattering integral $I(\mathbf{k})$ correspond to fixed points of the time evolution within the regime of applicability of the QBE \cite{Zakharov1992a}. 
Most prominent amongst these are the thermal fixed point corresponding to the system in thermal equilibrium and the trivial fixed point where the occupation number is independent of $\mathbf{k}$. 
At both fixed points the scattering integral vanishes and $n(\mathbf{k},t)$ becomes independent of $t$.
Note that both, the trivial distribution and the thermal Bose-Einstein distribution in the Rayleigh-Jeans regime, for $\omega(\mathbf{k})\sim k^{2}$, show a power-law behavior of the form \eq{scale} with $\zeta=0$ and $\zeta=2$, respectively.

The theory of weak wave turbulence \cite{Zakharov1992a} allows to analytically derive further, nonthermal fixed points at which the occupation number $n(\mathbf{k})$ obeys a scaling law of the form \eq{scale} and, in general, $\zeta\not=2$. 
As in classical turbulence of an incompressible fluid one assumes that universal scaling appears within a certain regime of momenta, the inertial range. 
According to this picture, outside the scaling regime kinetic energy enters the system from an external source and/or is dissipated into heat, whereas there are no sources and sinks within the inertial interval. 
Instead, kinetic energy is transported from momentum shell to momentum shell without loss or gain. 
To a good approximation, this process is described by a continuity equation in momentum space, with a momentum-independent, radially oriented current vector.\footnote{Note that justification of this assumption, i.e., locality of the transport, needs to be checked for each particular wave-turbulent solution \cite{Zakharov1992a}.}  
A central aspect of weak-wave-turbulence theory is that the QBE can be cast into different such equations~\cite{Zakharov1992a}, for the radial number density $N(k)=(2k)^{d-1}\pi n(k)$ and the energy density $E(k)=(2k)^{d-1}\pi \varepsilon(k)$, $\varepsilon(k)=\omega(k)n(k)$,
\begin{align}
  \partial_{t}N(k,t) 
  &= -\partial_{k}Q(k),
  \label{eq:BalEqQ}\\
  \partial_{t}E(k,t) 
  &= -\partial_{k}P(k).
  \label{eq:BalEqP}
  \end{align}
Depending on whether the radial particle current $Q(k)=(2k)^{d-1}\pi Q_{k}(k)$ or energy current $P(k)=(2k)^{d-1}\pi P_{k}(k)$ is taken to be independent of $k$, one derives different scaling exponents. 
The resulting exponents are
\begin{align}
   \zeta^\mathrm{UV}_{Q}= d-2/3 ,\quad
\zeta^\mathrm{UV}_{P}= d.
  \label{eq:kappaUV}
\end{align}
These exponents can be obtained by simple power counting:
Combining Eqs.~\eq{QKinEq} and  \eq{BalEqQ} gives the radial relation $\partial_{k}Q(k)\sim k^{d-1}I(k)$ which implies that stationarity requires $k^{d}I(k)$ to become $k$-independent, i.e. scale like $k^{0}$.
Counting all powers of $k$ in $I(k)$, \Eq{KinScattInt}, in the wave-kinetic regime where the terms of third order in the occupation numbers dominate the scattering integral, this requires  $n(k)\sim k^{-d+2/3}$.
Analogously one infers the exponent $\zeta^\mathrm{UV}_{P}$ from the balance equation \eq{BalEqP} for the energy density $\varepsilon(k)\sim k^{2}n(k)$.
Despite this simple procedure, the existence of the respective scaling solutions has to and can be derived rigorously from the QBE by means of Zakharov conformal integral transforms~\cite{Zakharov1992a}. 
We note that, similar to classical turbulence, the case $d=2$ is special, where the scaling exponent $\zeta_\mathrm{P}^\mathrm{UV}$ equals that of a thermal distribution in the Rayleigh-Jeans regime.

%-----------------------------------------------------------------------
\begin{figure}[t]
\begin{center} 
\includegraphics[width=0.35 \textwidth]{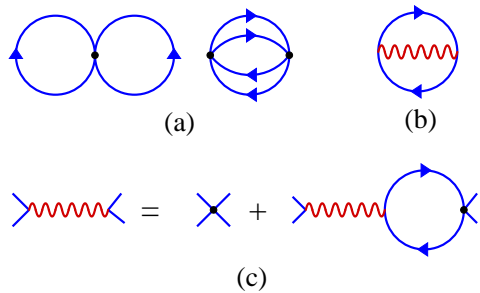}
\caption{(Color online) 2PI diagrams of the loop expansion of $\Gamma_2[G]$.
(a) The two lowest-order diagrams of the loop expansion which lead to the Quantum-Boltzmann equation. 
Black dots represent the bare vertex $\sim g\delta(x-y)$, solid lines the propagator $G(x,y)$.
(b) Diagram representing the resummation approximation which, in the IR, replaces the diagrams in (a) and gives rise to the scaling of the $T$-matrix in the IR regime.
(c) The wiggly line is the scalar propagator which is represented as a sum of bubble-chain diagrams.
See text for more details.
\label{fig:2PI}}
\end{center}
\end{figure}
%-----------------------------------------------------------------------

%======================================================================
\subsection{Strong wave turbulence}
\label{subsec:StrongWT}
Given a positive scaling exponent $\zeta$ momentum occupation numbers $n(k)\sim k^{-\zeta}$ grow large in the IR regime of small $k$.  
As a consequence, for a given coupling $g$, the QBE fails in this regime, where contributions to the scattering integral $I(k)$ which are of higher order than   $g^{2}n^{3}$ become important. 

To find scaling solutions in the IR, an approach beyond kinetic theory is required. 
This is available through quantum-field dynamic equations derived from the two-particle irreducible (2PI) effective action or $\Phi$-functional \cite{Luttinger1960a, Baym1962a, Cornwall1974a} beyond the 2-loop order in the expansion of the self-energy.
See Ref.~\cite{Scheppach:2009wu} for details of the procedure summarized in the following.
The 2PI equations include the Dyson equation for the time-ordered Green function $G(x,y)=\langle\mathcal{T}\Phi^{\dagger}(x)\Phi(y)\rangle$ (here, we use four-vector notation $x=(x_{0},\mathbf{x})$), from which a time evolution equation \eq{QKinEq} for $n(\mathbf{k})$ is derived. 
As before one considers zeros of the scattering integral which in the dynamic theory reads
\begin{equation}
  I(\mathbf{k}) = \int d\omega\,[\Sigma^\rho(k)F(k) - \Sigma^F(k)\rho(k)].
\label{eq:stationarity}
\end{equation}
Here, $k\equiv(\omega,\mathbf{k})$, and $\rho$ and $F$ are the spectral and statistical components of $G$, respectively, defined in coordinate space by $F(x,y)=\langle\{\Phi^{\dagger}(x),\Phi(y)\}\rangle/2$, $\rho(x,y)=i\langle[\Phi^{\dagger}(x),\Phi(y)]\rangle$,  $G(x,y) = F(x,y) -({i}/{2})\mathrm{sgn}(x_0-y_0)\rho(x,y)$.
The corresponding contributions to the self energy $\Sigma(x,y)=2i\delta\Gamma_{2}/\delta G(x,y)$ are defined in terms of $G$ through a loop expansion of the 2PI effective action,  see \Fig{2PI}.
Resumming an infinite set of such loop diagrams contributing to the 2PI effective action \cite{Berges:2001fi,Aarts:2002dj} leads to a non-perturbative, effectively renormalized coupling in the dynamic equations \cite{Berges:2008wm,Scheppach:2009wu}.

To derive the scaling solutions of the dynamic equations for Green functions $G(\omega,\mathbf{k})$ one assumes separate scaling of its spectral and statistical components according to $\rho(s^z\omega,s\vec k) =  s^{-2+\eta}\rho(\omega,\vec k)$,  $F(s^z\omega,s\vec k) =  s^{-2-\kappa}F(\omega,\vec k)$, $s>0$.
Here, $z$ is the dynamical scaling exponent accounting for a different scaling in $\omega$ as compared to $\mathbf{k}$.
The scaling exponent $\zeta$ is related to $\kappa$ by $n(s\vec k) = s^{z-2-\kappa}n(\vec k)$.
$\kappa$ is derived from the condition that the scattering integral \eq{stationarity} vanishes, making use of Zakharov transformations,
while the anomalous scaling exponent $\eta$ remains undetermined by this and, for the first, can be assumed to vanish for the $d=2$ and $3$ dimensional cases considered in this paper. 
The IR scaling exponents for radial quasiparticle flow ($Q$) and radial energy flow ($P$) in $d$ dimensions were predicted in Ref.~\cite{Scheppach:2009wu} to be
\begin{align}
   \zeta^\mathrm{IR}_{Q}= d+2 ,\quad
\zeta^\mathrm{IR}_{P}= d+2+z \,.
  \label{eq:kappaIR}
\end{align}

In situations where a quasiparticle picture applies the dynamical exponent $z$ corresponds to the homogeneity index of $\omega$: $\omega(s\mathbf{k}) = s^{z}\omega(\mathbf{k})$.
In this case the nonperturbatively resummed effective coupling can be related to the diagonal elements of an effective many-body $T$-matrix, $T^{\mathrm{eff}}_{\vec k\vec p\vec q\vec r}\equiv T^{\mathrm{eff}}_{\vec k+\vec p,\vec q+\vec r}$ in the kinetic Boltzmann formulation. 
In the scaling regimes this scales as  $|T^{\mathrm{eff}}_{\mathbf{k}}|\equiv|T^{\mathrm{eff}}_{\mathbf{k},\mathbf{k}}|\sim |gCk^{z-2}/[1+C'gk^{d-2}n_{\mathbf{k}}]|$, $k=|\mathbf{k}|$, where $C'$ is some constant which fine-tunes the position of the transition from UV to IR scaling:
For small $n_{k}$ and $z=2$ one recovers the UV case discussed in the previous section, i.e., $T^{\mathrm{eff}}_{\mathbf{k}}$ is a constant independent of $k$.
For large $n_{k}$, the second term in the denominator dominates which, assuming scaling of $n_{\mathbf{k}}\sim k^{-\zeta}$, implies a power-law behavior $|T^{\mathrm{eff}}_{\mathbf{k}}|^{2}\sim k^{2(\zeta-d+z)}$ and, in turn, the modified scaling \eq{kappaIR} of $n_{\mathbf{k}}$ in the infrared regime of small wave numbers as compared to the UV regime discussed before. 
Physically, the renormalized $T$-matrix implies a reduction of the effective interaction strength in the IR regime of strongly occupied modes \cite{Berges:2008wm}. 
As a consequence, single-particle occupation numbers rise, towards smaller wave numbers, in an even steeper way than in the weak-turbulence regime.

%======================================================================
 \begin{figure}
\center
 \includegraphics[width=0.4\textwidth]{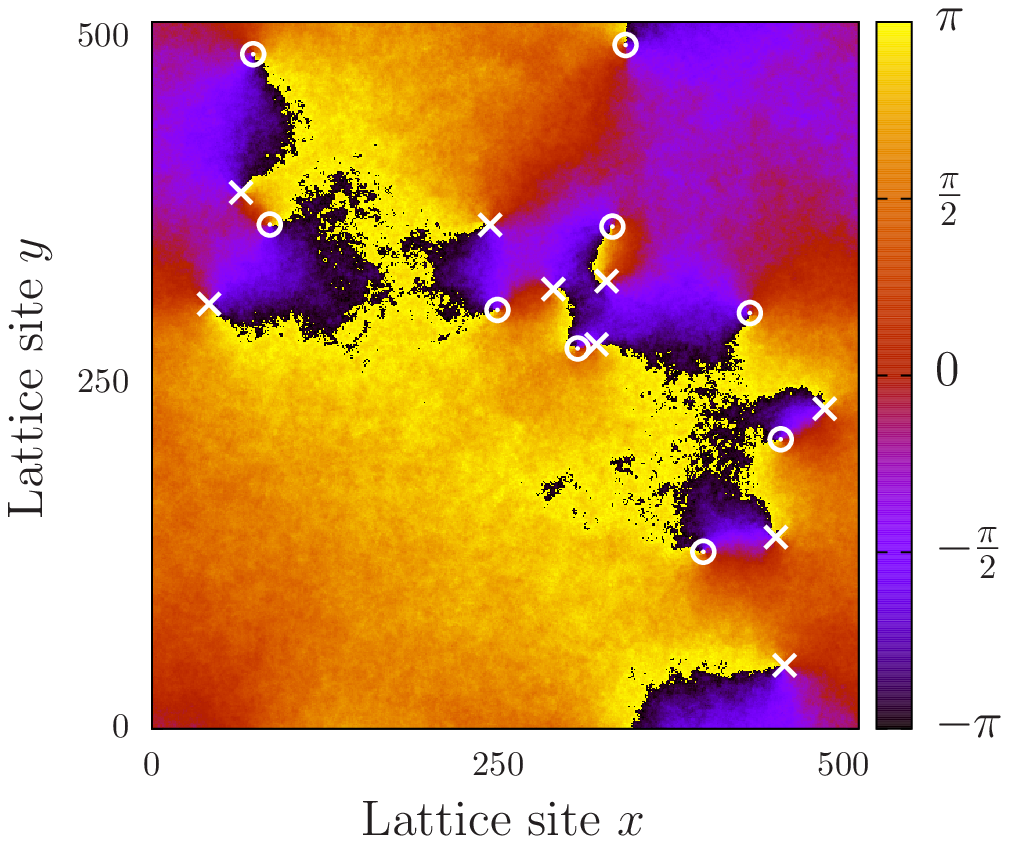}
 \caption{(Color online) The plot shows the phase of the complex field for a single run of the evolution in $d=2$ dimensions. 
 Parameters: $\overline{g}=3\times 10^{-5}$, $N=4\times10^8$, $N_s=512$, $\overline{t}=46340$, coordinates in lattice units of $a_{s}$, see~\App{classfields}.
 White rings (crosses) mark vortices (antivortices). For a video of the evolution see~\cite{videos}.}
\label{fig:2DPhase512}
\vspace{0.5cm}
 \includegraphics[width=0.45\textwidth]{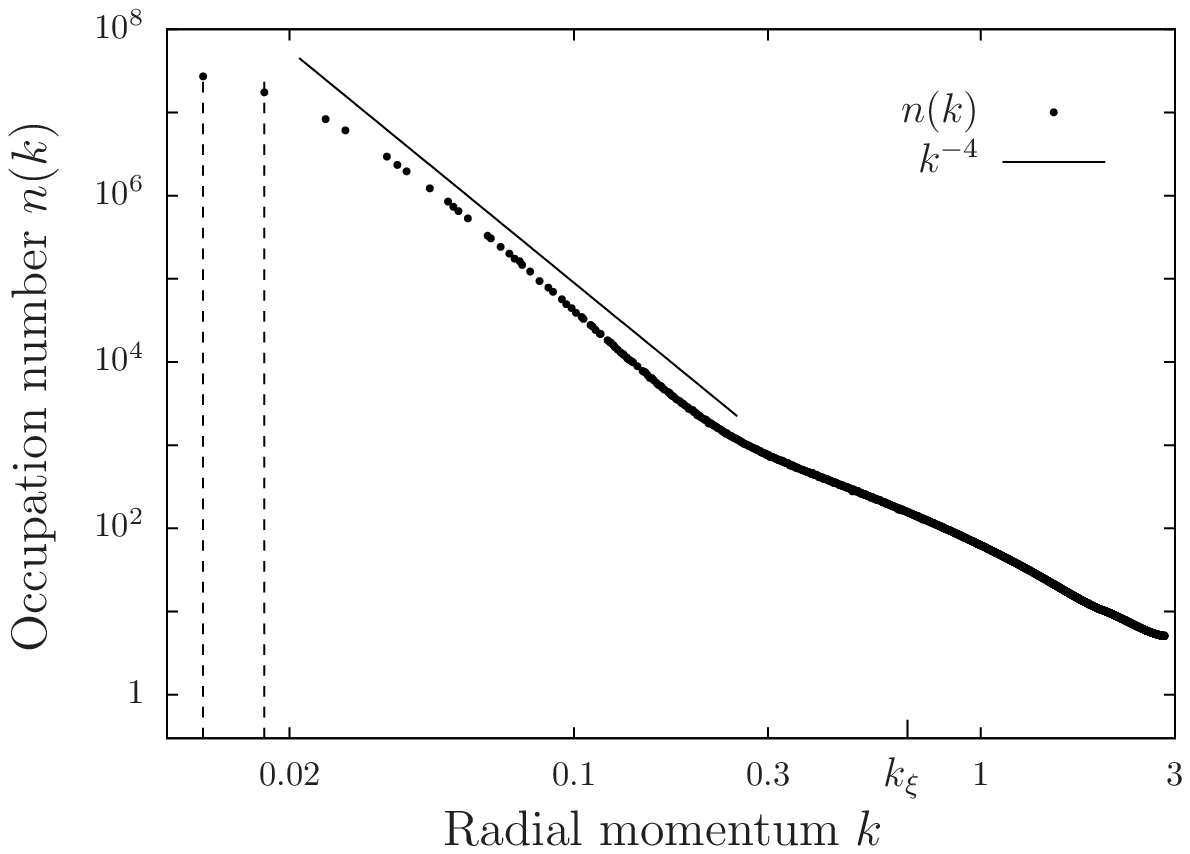}
 \caption{Single-particle mode occupation numbers are shown as functions of the radial momentum $k$ at the time shown in \Fig{2DPhase512}. 
 $k$ is in lattice units, see \App{classfields}. 
 Note the double-logarithmic scale. 
 The development of a IR scaling with $n(k)\sim k^{-4}$ coincides with the presence of superfluid vortices, with additional wave-turbulent or thermal background short-wavelength fluctuations in the UV. Dashed lines indicate the filling of the initially occupied modes with $k>0$.}
\label{fig:2DModeOccupation}
 \end{figure}
%======================================================================
%
%
%======================================================================
 \begin{figure}[!h]
\center
 \includegraphics[width=0.41\textwidth]{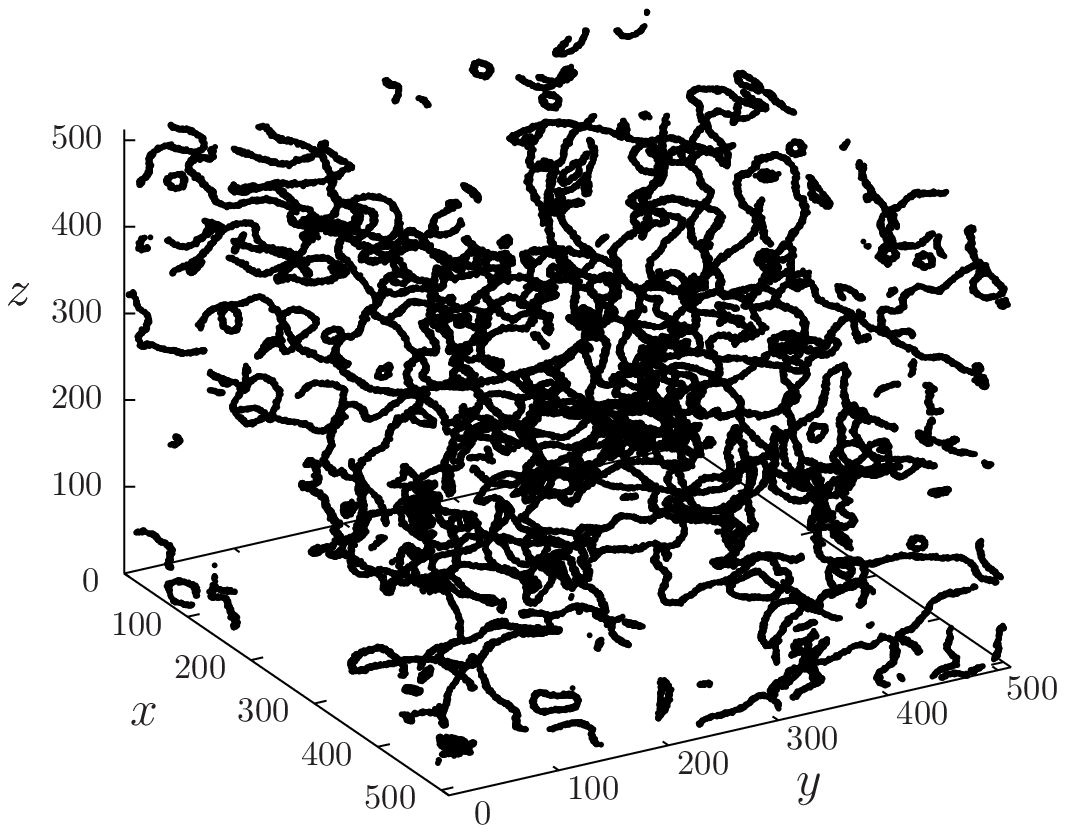}
 \caption{Vortex line tangles are shown in three dimensions. 
 Black dots indicate where the amplitude of the complex field falls below 5\% of the mean density $n$. 
 Parameters are:  $\overline{g}=4\times 10^{-4}$, $N= 6.4 \times 10^{10}$, $N_s=512$, $\overline{t}=3276$, coordinates in lattice units, see~\App{classfields}. For a video of the evolution see~\cite{videos}.}
\label{fig:Phase512}
\vspace{0.6cm}
 \includegraphics[width=0.45\textwidth]{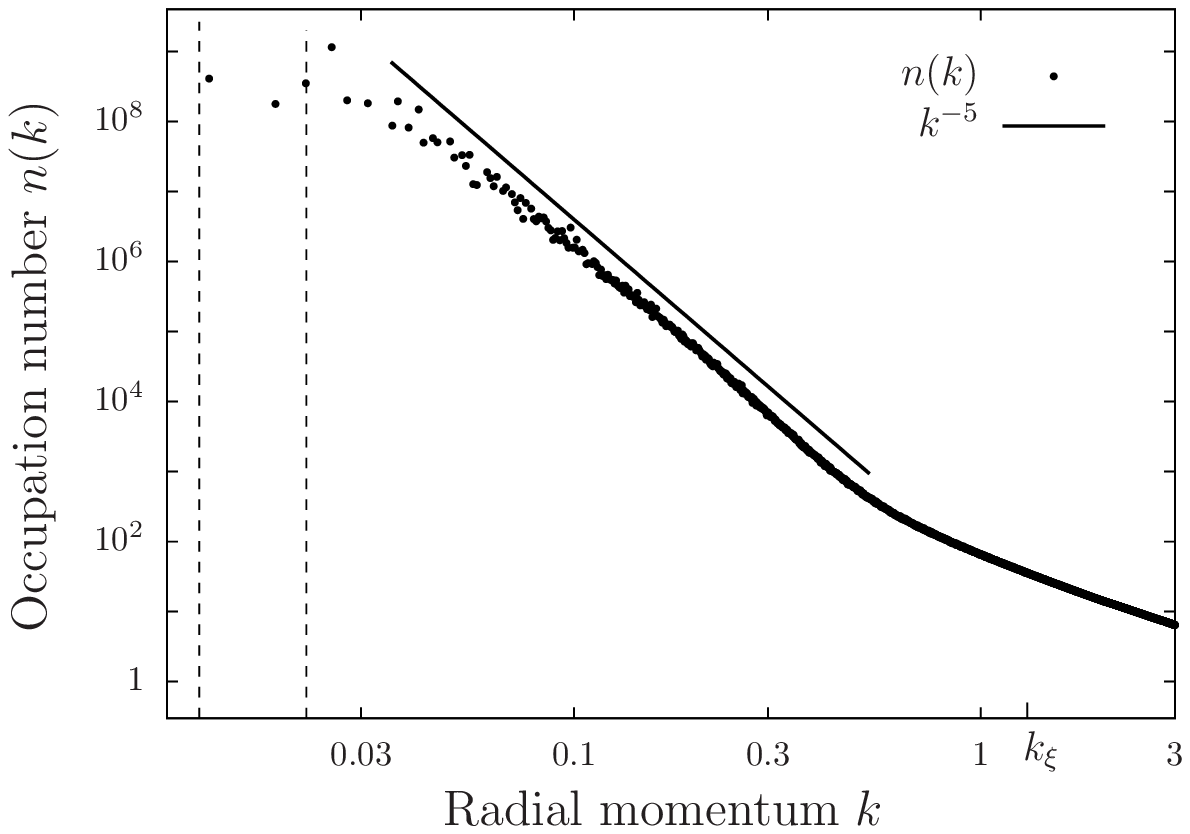}
 \caption{Single-particle mode occupation numbers are shown as functions of the radial momentum $k$ for the snapshot in \Fig{Phase512}. 
 $k$ is in lattice units. 
 Note the double-logarithmic scale. The development of a IR scaling $n(k)\sim k^{-5}$ coincides with the presence of superfluid vortex lines.
 Dashed lines indicate the filling of the initially occupied modes with $k>0$.}
\label{fig:ModeOccupation3g8}
 \end{figure}
%======================================================================

%======================================================================
\subsection{Wave-turbulent scaling and vortices \newline in a Bose gas}
\label{subsec:SummaryVortexNumerics}
We now briefly review the relation between wave-turbulent scaling and the appearance of vortical excitations in an ultracold degenerate Bose gas. 
In \cite{Nowak:2010tm} the results of semiclassical simulations of the gas dynamics were reported, obtained by solving the classical field equation
\begin{equation}  \label{eq:GPE}
   \i \del_t \phi(\mathbf{x},t)= \left[ -\frac{\nabla^2}{2m}+g|\phi(\mathbf{x},t)|^2 \right] \phi(\mathbf{x},t) 
\end{equation}
in a box with periodic boundary conditions. See \Appendix{classfields} for details on the simulations and on lattice units.
The initial field $\phi(\mathbf{x},0)$ was prepared by macroscopically populating a few of the lowest momentum modes in the computation such that the resulting condensate density in configuration space varied between zero and some maximum value. Quantum noise is taken into account by adding a small random contribution to each field mode.
Vortical excitations were created in large numbers, within shock-waves forming during the non-linear evolution of the coherent matter-wave field. 
\Fig{2DPhase512} shows an example of the condensate phase distribution over a numerical lattice in $d=2$ dimensions, with vortices and antivortices indicated by white crosses and circles, respectively. 
\Fig{Phase512} depicts the location of vortex-ring cores in a $d=3$-dimensional system. For videos of the evolution see~\cite{videos}.
Both figures show a snapshot of the system at some time after the vortical excitations have formed. 
At these times, in $d=2$, part of the vortex-antivortex pairs have re-annihilated with each other already, and in $d=3$ some of the rings have undergone reconnections and also disappeared by shrinking to zero size. 

Runs were repeated many times ($\sim 100$ times in $d=2$ and $\sim 10$ times in $d=3$) for an ensemble of initial configurations differing by statistical noise. The number of runs was chosen such that the statistical error arising from run-to-run fluctuations was reduced to a value on the order of the size of the symbols used in the figures.
In the regime of momenta with large occupation numbers where quantum statistical fluctuations play little role, correlation functions like the spectrum $n_{k}$ could be computed in a quasi exact way by averaging at a given time over the ensemble. 
In this way, the flow was analysed in terms of the ensemble- and angle-averaged single-particle spectrum  
\begin{equation}\label{eq:spectrumdef}
n(k) = \int \mathrm{d}^{d-1}\Omega_{k} \, \langle \phi^*(\mathbf{k})\phi(\mathbf{k}) \rangle_{\mathrm{ensemble}},
\end{equation}
as a function of radial momentum $k=|\mathbf{k}|$. Figs.~\fig{2DModeOccupation} and \fig{ModeOccupation3g8} show this spectrum for  $d=2$ and $3$, at the times chosen in Figs.~\fig{2DPhase512} and \fig{Phase512}, respectively. 
Most importantly, these spectra show scaling, \Eq{scale}, within a range of momenta between about the maximum of the initially occupied momentum modes indicated by the vertical lines and $k\simeq 0.3$ in lattice units. 
The exponent $\zeta$ is in agreement with the field theoretical prediction $\zeta_\mathrm{Q}^\mathrm{IR}=d+2$ given in \Eq{kappaUV}.

It was found that the annihilation processes which eventually destroy the vortical structure while coherence builds up in the system are very slow and thus stabilize the scaling solution over a long time. In particular, only after the last vortex-antivortex pair has annihilated and the last ring shrunk to zero the power law $n(k)\sim k^{-\zeta_\mathrm{Q}^\mathrm{IR}}$ breaks down and a thermal distribution of particles is left over. Characteristic times are $\overline{t}\sim \mathcal{O}(10^2)$ for vortex formation, $\overline{t}\sim \mathcal{O}(10^3)$ for stabilization of the scaling solution and $\overline{t}\sim \mathcal{O}(10^4)-\mathcal{O}(10^5)$ for the last vortices to annihilate.

%
%== Point vortex model ================================================================
\section{Point- and line-vortex models}
\label{sec:RandomVortexModel}
\begin{figure}
 \includegraphics[width=0.20\textwidth]{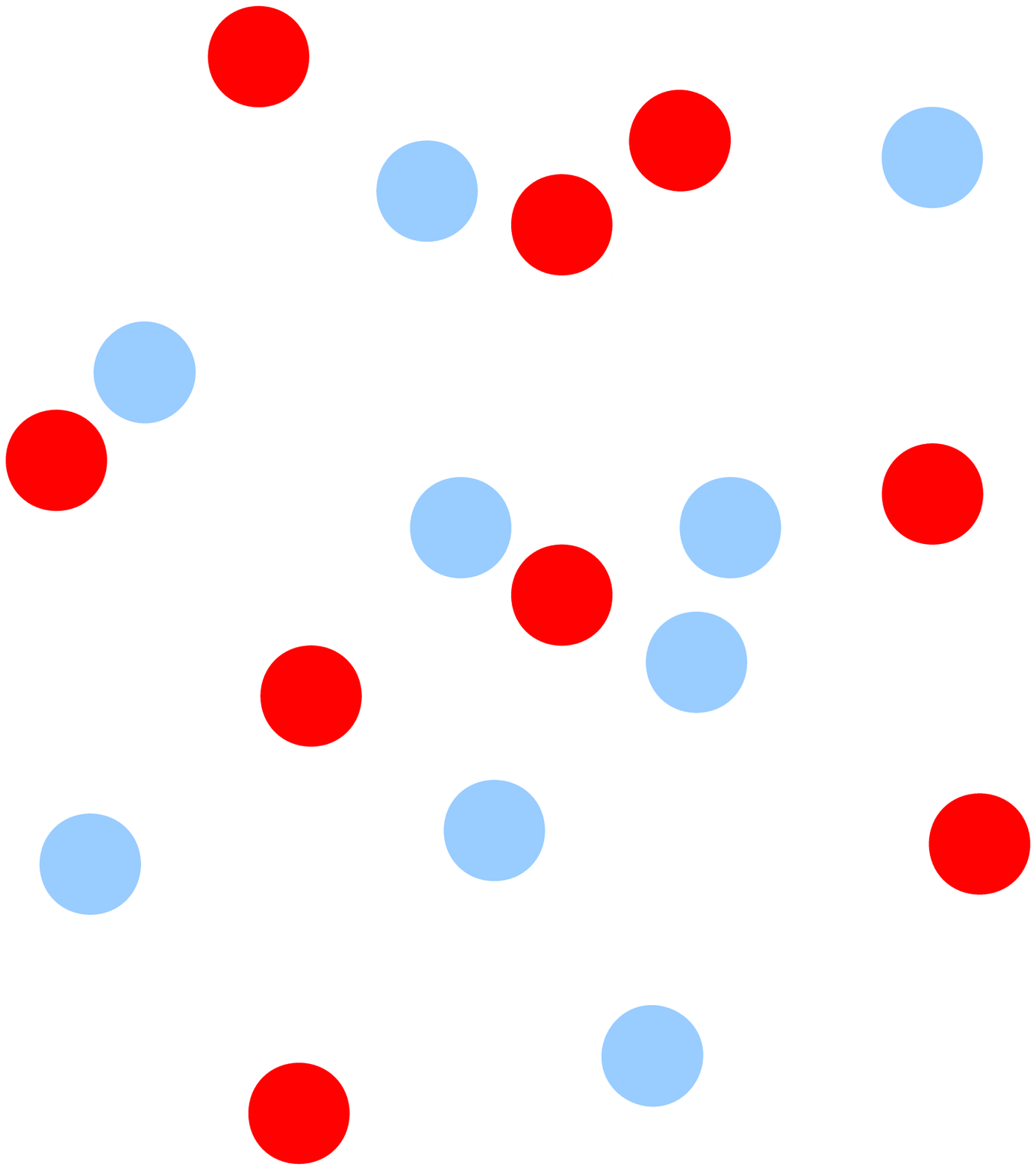}
\hspace{0.02\textwidth}
 \includegraphics[width=0.20\textwidth]{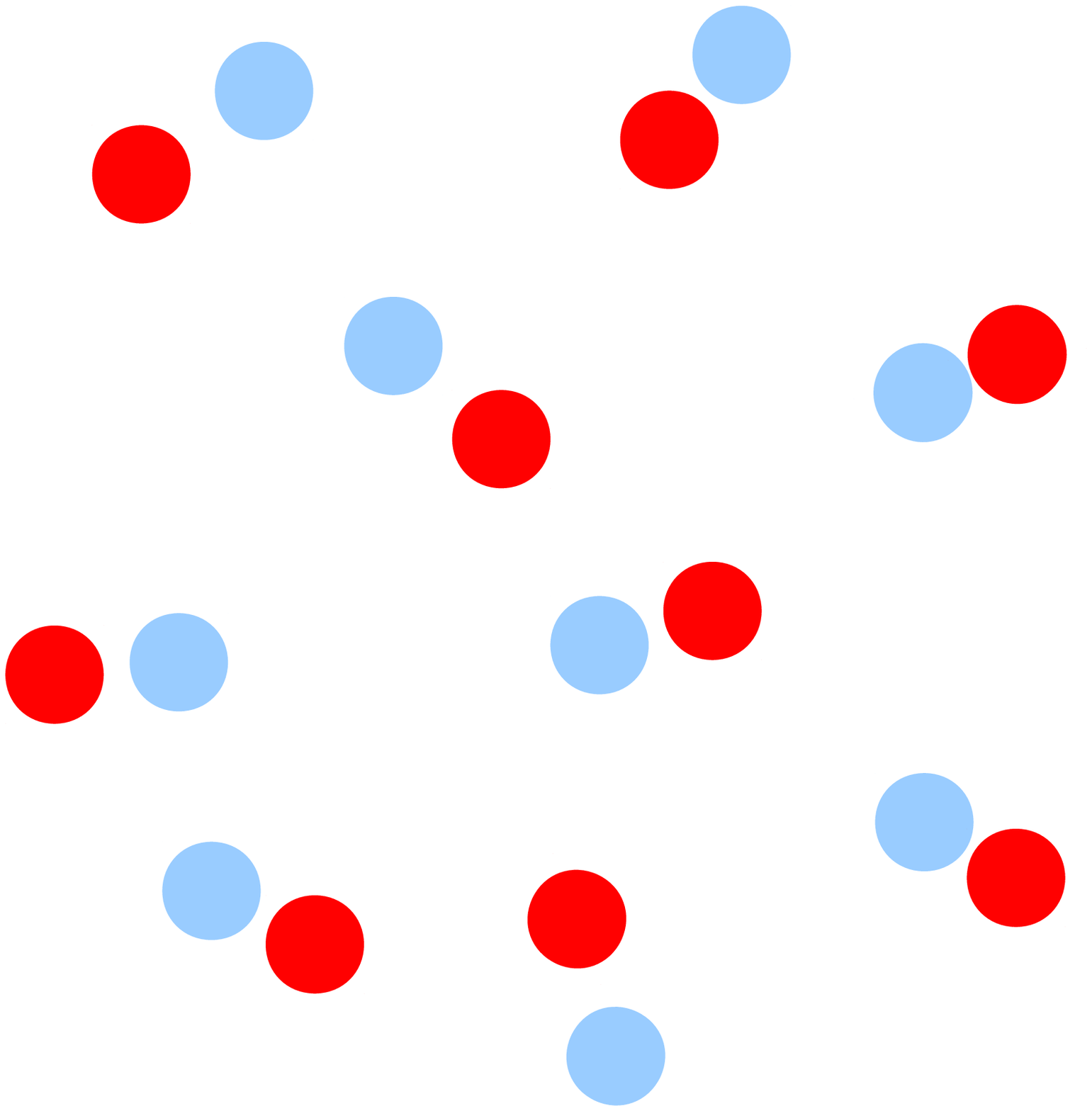}
\vspace{-0.08\textwidth}
\caption{(Color online) Left panel: Sketch of a random vortex-antivortex distribution underlying the IR scaling close to the nonthermal fixed point. 
Right panel: Correlated vortex distribution causing a modification to weaker pair scaling in the IR, for momenta smaller than the inverse of the average pairing length.}
 \label{fig:VortexSystemRandom}
 \end{figure}

As we have seen above, the scaling of the momentum occupation numbers in the IR regime of large wave numbers, predicted within a non-perturbative analysis of strong wave-turbulence, correspond, for $2$- and $3$-dimensional degenerate Bose gases, to the appearance of macroscopic vortical excitations. 
In the following, we analyse the observed scaling spectra by comparing them to the single-particle momentum distributions for a set of randomly positioned point vortices (vortex lines) in $d=2$ ($d=3$) dimensions.
We will show that uncorrelated vortices (vortex lines) are sufficient to yield the infrared scaling with exponent $\zeta_\mathrm{Q}^\mathrm{IR}$, see \Eq{kappaIR}. 
We will also show that beyond this, pair correlations between vortices and antivortices as well as configurations with small rings well separated from each other can give rise to a further scaling exponent deviating from $\zeta_\mathrm{Q}^\mathrm{IR}$ (see \Fig{VortexSystemRandom}). 

The point (line) vortex model employed here was introduced by Onsager in 1949 \cite{Onsager1949a}.
It describes the complex flow pattern in terms of the statistical mechanics of interacting classical point objects. This model has been constructed as a discrete vorticity approximation of classical fluid turbulence, but it is even more suitable to describe superfluid turbulence consisting of quantized vortices. Further applications include plasma physics or stellar dynamics \cite{Eyink2006a,Chavanis2002a,Berdichevsky1995a}.

In the following, we study the spectrum of a set of vortices, finding different scaling regimes dependent on vortex correlation functions. 
With this approach, nonthermal fixed points in an ultracold Bose gas can be related to the statistics of vortices. 
Numerical calculations, sampling field configurations of static randomly positioned vortices, confirm the analytical result. 
Finally, the velocity probability distribution of a vortex-dominated flow ~\cite{White2010a,Paoletti2008a} is derived and numerically confirmed.
%

%======================================================================
\subsection{Independent vortices in $d=2$}
\label{subsec:IndependentVortices}
%
%
%======================================================================
%======================================================================
%
%
In two dimensions, an isolated, singly quantized vortex is described by the complex field $\phi(r,\varphi) \equiv \sqrt{n(r)} e^{i\varphi}$. 
As the $r$-dependence of the density $n(r)$ only becomes important at small scales on the order of the healing length $~\xi=1/\sqrt{2mgn}$ where our simulations are dominated by thermal excitations, we will omit it in the following and assume $n$ to be uniform.

A system of $M$ vortices in $d=2$ dimensions can be defined as $\phi(\mathbf{x}) = \Pi_i^M \phi_i(\mathbf{x})$, where $\phi_i(\mathbf{x}) = \phi(\mathbf{x}-\mathbf{x}_i)$ is the single vortex field centered around $\mathbf{x}_i$. We derive the corresponding particle spectrum by considering the hydrodynamic velocity field $\mathbf{v} =  \nabla \varphi/m$. Denoting the velocity field of a single vortex as $\tilde{\mathbf{v}}(\mathbf{x})$ (see \Appendix{SingleVortex}) we can express the mean classical kinetic energy density of the velocity field as 
\begin{eqnarray}  \label{eq:EnergySpec}
E_\mathrm{v}(\mathbf{x}) 
&=& \frac{m}{2}  \, \langle |\,  \mathbf{v}(\mathbf{x}) |^2 \rangle 
\nonumber \\
&=& \frac{m}{2} \, \langle |\, \int \mathrm{d}^2x' \,  \tilde{\mathbf{v}}(\mathbf{x} - \mathbf{x'}) \, \rho(\mathbf{x'}) \, \, |^2 \rangle \, ,
\end{eqnarray}
where $\rho(\mathbf{x}) = \sum_{i=1}^{M} \kappa_i \delta(\mathbf{x}-\mathbf{x}_i)$ defines the spatial distribution of vortices with winding number $\kappa_i = \pm 1$. Here and in the following, $\langle\cdot\rangle$ denotes an ensemble average over different realizations of the classical field $\phi(\mathbf{x})$.
We derive the low-$k$ scaling of $n(k)$ from the kinetic-energy spectrum $E_\mathbf{v}(k)$, given by the angle-averaged Fourier transform of $E_\mathbf{v}(\mathbf{x})$,  taking into account that at low $k$,  the single-particle spectrum is dominated by the superfluid velocity field $\mathbf{v}$, i.e., 
\begin{equation}  \label{eq:nkfromEvk}
  n(k) \simeq 2mk^{-2} E_\mathbf{v}(k).
\end{equation}
One has, from  \Eq{EnergySpec}
\begin{equation}  \label{eq:Evk}
 E_\mathrm{v}(\mathbf{k}) \sim \langle |\mathbf{v}(\mathbf{k})|^2 \rangle = \langle  \, |\rho(\mathbf{k})|^2  \, |\tilde{\mathbf{v}}(\mathbf{k})|^2 \rangle \,,
\end{equation}
with 
\begin{equation}  \label{eq:Vortexdensity}
|\rho(\mathbf{k})|^2  =  \sum_{i,j}^M  \kappa_i \kappa_j e^{i\mathbf{k}(\mathbf{x}_i-\mathbf{x}_j)} \,.
\end{equation}
Below the healing length scale $ k_{\xi} = 2\,\mathrm{sin}(\pi/2\xi) $ (lattice units), the modulus of the velocity field of a single vortex scales as $ | \tilde{\mathbf{v}} | \sim k^{-1} $ and is radially symmetric in momentum space. Hence, the angle-averaged single-particle spectrum scales like
\begin{equation}  
\label{eq:VE}
n(k) = k^{-4}  \left( M + 2 \sum_{i<j} \kappa_i \kappa_j J_0( kl_{ij} ) \right) .
\end{equation}
Here, $J_0(y) = (2\pi)^{-1} \int_{-\pi}^{\pi} \mathrm{d}\theta \, \mathrm{cos}(y\,\mathrm{cos}(\theta))$ denotes the zeroth-order Bessel function and $l_{ij}=|\mathbf{x}_i-\mathbf{x}_j|$ is the distance between vortices $i$ and $j$.

Assuming that the positions $\mathbf{x}_{i}$ of the vortices are uncorrelated one can take the average over relative positions $l_{ij}$ within the area $V_{R}=\pi R^{2}$,
\begin{equation}  \label{eq:AV}
 \frac{2\pi}{V_R} \int_0^R \mathrm{d}l \,l\, J_0(kl)  =  2\frac{J_1(kR)}{kR},
\end{equation}
and, for fixed $k$, the limit $R \rightarrow \infty$.
Hence, the second term in brackets in \Eq{VE} vanishes and one finally obtains the scaling~\cite{Novikov1976a}
\begin{equation}
 n(k)\sim k^{-4}\,.
\end{equation}
%
%

%======================================================================
%
\subsection{Independent vortex-antivortex pairs}
\label{subsec:IndPairs}
In a two-dimensional superfluid containing vortices and antivortices an effectively attractive force between the two species can lead to pair correlations. We study a signature of this feature in the single-particle momentum spectrum by applying the point vortex model introduced above to the case of vortex-antivortex pairs. 

As a first step we calculate the velocity field $\mathbf{v}_{\mathrm{VA}}$ for a vortex-antivortex pair with the vortex situated at $\mathbf{x}_{1}$ and the antivortex at $-\mathbf{x}_{1}$, 
\begin{equation}  \label{eq:vVA}
\mathbf{v}_{\mathrm{VA}} =  \tilde{\mathbf{v}}(\mathbf{x}-\mathbf{x}_1) - \tilde{\mathbf{v}}(\mathbf{x}+\mathbf{x}_1) .
\end{equation}
The squared velocity field far away from the center of the pair can be obtained via a dipole approximation $|\mathbf{x}| \gg |\mathbf{x}_1|$ which yields the scaling $\mathbf{v}_{\mathrm{VA}}\sim r^{-2}$. 
Hence, in Fourier space, the pair velocity field scales as $|\mathbf{v}_{\mathrm{VA}}|^2 \sim k^{0}$ for low momenta. In this regime, the vortex-antivortex pair can again be treated as a point-like object with modified velocity scaling. To obtain the infrared scaling of a set of random vortex-antivortex pairs, we define a spatial pair distribution $\rho_\mathrm{pair}(\mathbf{x}) = \sum_i \delta(\mathbf{x}-\mathbf{x}_i)$, with $\mathbf{x}_i$ denoting the center of the $i$-th vortex-antivortex pair. Then, the analysis performed for random vortices above can be adopted. Therefore, in the case of independently distributed pairs, the approach predicts the infrared scaling of the occupation number to be the same as for a single pair, i.e. $n(k) \sim k^{-2}$. 

%======================================================================
%
\subsection{Pair correlated vortices in $d=2$}
\label{subsec:CorrelatedVortices}

The considerations from \Subsect{IndPairs} are only valid in the far IR (or equivalently for pair size going to zero).
In our numerical simulations we found, however, that for large evolution times, a finite minimum distance emerged between vortices and antivortices, and also between vortices with the same circulation, see \Subsect{PairingSpectrum}. 
 To take into account this observation and to analyse the full spectrum, we go back to writing the distribution $\rho(\mathbf{x}) = \rho^\v(\mathbf{x}) - \rho^\a(\mathbf{x})$ as the sum of distributions $\rho^{\v}(\mathbf{x})= \sum_{i=1}^{M} \delta( \mathbf{x}-\mathbf{x}_i^\v)$ of $M$ vortices and $\rho^\a(\mathbf{x})= \sum_{i=1}^{M} \delta(\mathbf{x}-\mathbf{x}_i^\a)$ of $M$ antivortices. 
Hence,
\begin{eqnarray}  \label{eq:CF}
\langle \, |\rho(\mathbf{k})|^2 \, \rangle  = \int \mathrm{d}^2x \, \mathrm{d}^2x' \, e^{i\mathbf{k}(\mathbf{x}-\mathbf{x'})} C(\mathbf{x},\mathbf{x}') \,,
\end{eqnarray}
with $C(\mathbf{x},\mathbf{x}')= \langle \, \rho^\v_\mathbf{x} \rho^\v_\mathbf{x'} \, \rangle - \langle \,\rho^\v_\mathbf{x} \rho^\a_\mathbf{x'}  \, \rangle- \langle \,\rho^\a_\mathbf{x} \rho^\v_\mathbf{x'}  \, \rangle + \langle \,\rho^\a_\mathbf{x} \rho^\a_\mathbf{x'} \, \rangle$.
This allows for a derivation of the kinetic-energy distribution in terms of correlation functions of vortex positions. 

We model pairing by the density-density correlation functions
\begin{eqnarray}
\langle  \rho^{\v(\a)}_{\mathbf{x}} \rho^{\v(\a)}_{\mathbf{x'}} \rangle 
&=& \frac{M}{V_R}\delta(\mathbf{x}-\mathbf{x'}) + P_{\mathbf{x},\mathbf{x}'},  
\label{eq:rhoVrhoV}
\\
\langle  \rho^{\v(\a)}_{\mathbf{x}} \rho^{\a(\v)}_{\mathbf{x'}} \rangle 
&=& \frac{M}{V_RV_\lambda}\theta(\lambda - |\mathbf{x}-\mathbf{x'}| ) +  P_{\mathbf{x},\mathbf{x}'}.  
\label{eq:rhoVrhoA}
\end{eqnarray}
where $V_\lambda= \pi \lambda^2$ is the area in which the theta function equals one.  
The contributions
\begin{eqnarray}
P_{\mathbf{x},\mathbf{x}'} &=&  \frac{M\left({M}-1 \right)}{V_{R}(V_{R}-V_{\Lambda})}\theta(|\mathbf{x}-\mathbf{x'}| -\Lambda) 
\end{eqnarray}
take into account that, besides the pairing, vortices and antivortices avoid each other in the dilute gas, keeping a minimum distance $\Lambda$. This is due to vortex-vortex repulsion and fast vortex-antivortex annihilation on small distances. 
The functions $P_{\mathbf{x},\mathbf{x}'}$ cancel out in \Eq{CF}.\footnote{If different avoidance scales $\Lambda$ apply for vortices and antivortices, the terms do not cancel, but the remaining term does not alter the results for pair scaling derived here.}

Inserting Eqs.~\eq{rhoVrhoV}, \eq{rhoVrhoA} into \Eq{CF}, evaluating the Fourier transform and angular averaging gives 
\begin{eqnarray}  \label{eq:RHOSQUARE}
\langle \, |\rho(k)|^2 \, \rangle = 2M \left( 1-\frac{2}{k\lambda} J_1(k\lambda) \right) \,. 
\end{eqnarray}
The expansion of the integral for $ k\ll 2\pi / \lambda$ yields the leading-order result
\begin{equation}  \label{eq:rhoksqPairs2d}
\langle \, |\rho(k)|^2 \, \rangle = {M(k\lambda)^2}/{4}+\mathcal{O}(k^{4})
\end{equation}
and, from Eqs.~\eq{nkfromEvk} and \eq{Evk}, the same occupation-number scaling as for independent pairs, 
\begin{equation}  \label{eq:PS}
n(k) \sim k^{-2} \,.
\end{equation}
At momenta $ k\gg 2\pi / \lambda$ the independent-vortex scaling $n(k)\sim k^{-4}$ is restored. 
We note that the infrared result \eq{PS} can also be achieved by choosing more general pair correlations of the form $\langle  \rho^\v_\mathbf{x}\rho^\a_\mathbf{x'} \rangle = \langle \rho^\a_\mathbf{x}\rho^\v_\mathbf{x'}\rangle = {M\epsilon}[2V_R\pi(\lambda_{\mathrm{max}}^\epsilon-\lambda_{\mathrm{min}}^\epsilon)]^{-1}|\mathbf{x}-\mathbf{x'}|^{\epsilon-2}\theta(\lambda_{\mathrm{max}} - |\mathbf{x}-\mathbf{x'}| )\theta(  |\mathbf{x}-\mathbf{x'}| -\lambda_{\mathrm{min}}) +  P_{\mathbf{x},\mathbf{x}'}$, with $\epsilon\ge 0$ and $\lambda_{\mathrm{min}}<\lambda_{\mathrm{max}}$, which is closer to our numerical observations and includes the case of a fixed ``binding length'' $\lambda_{\mathrm{min}}\to\lambda_{\mathrm{max}}$. 
As before, pair scaling $k^{-2}$ is found for $k\lambda_{\mathrm{max}}\ll 1$, critical scaling $k^{-4}$ for $k\lambda_{\mathrm{max}}\gg 1$, irrespective of $\epsilon$.
%
%
%======================================================================
 \begin{figure}
 \includegraphics[width=0.45\textwidth]{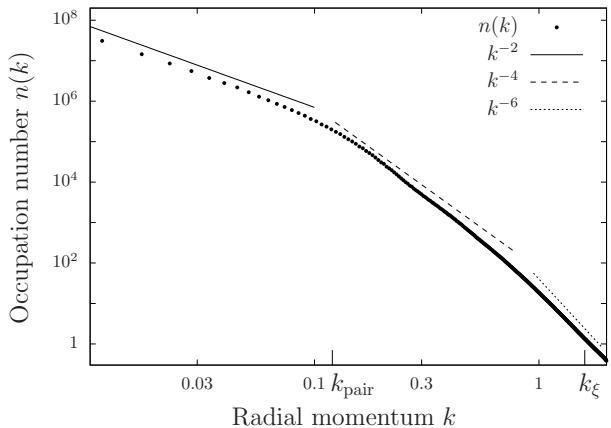}
 \caption{Radial momentum distribution for a randomly distributed set of $M=80$ bound vortex-antivortex pairs, on a  $N_s^{2}=1024^{2}$ grid. Note the double-logarithmic scale. The vortices are positioned according to the probability distribution \eq{PD}.
  At low momenta the power law is consistent with the existence of random vortex pairs $ n_k \sim k^{-2} $. 
Above $k_{\mathrm{pair}}$, the distribution exhibits the scaling of an ensemble of independent vortices, $n_k \sim k^{-4}$, while at momenta larger than the healing-length scale $k_\xi$ one observes the vortex-core scaling $\sim k^{-6}$. 
}
 \label{fig:RandomVortex}
\end{figure}
%======================================================================
%
We supplement our discussion of scaling in the point vortex model with numerical data obtained by averaging over an ensemble of field configurations in $d=2$ dimensions.
These configurations were constructed by multiplying uncorrelated single-vortex fields centered at positions according to the probability distribution 
\begin{eqnarray}\label{eq:PD}
P^{(M)}(\mathbf{x}^\v_1, \mathbf{x}^\a_1, ... ) = \prod_i^{M} P^{(1)}(\mathbf{x}^\v_i)P^{(2)}( \mathbf{x}^\v_i, \mathbf{x}^\a_i)
\end{eqnarray}
with $ P^{(1)}(\mathbf{x}^\v_i) = {1}/{V_R}, P^{(2)}( \mathbf{x}^\v_i, \mathbf{x}^\a_i) = V_\lambda^{-1} \theta (\lambda - |\mathbf{x}^\v_i - \mathbf{x}^\a_i|) $, where we neglect that unpaired vortices avoid each other, i.e., choose $\Lambda=0$.
They do not represent stable solutions of the classical field equation nor do they contain sound-wave and related excitations yet which would build up through the interactions of the field and the vortices.

The resulting momentum spectrum is presented in \Fig{RandomVortex}. Three scaling regimes can be observed. At low momenta the power law is consistent with the existence of random vortex pairs $ n_k \sim k^{-2} $. 
Above $k_{\mathrm{pair}}=2\mathrm{sin}\,(\pi/2\lambda)\simeq \pi/\lambda $, choosing $\lambda=25$ (lattice units), the distribution exhibits the scaling of an ensemble of independent vortices, $n_k \sim k^{-4}$, while at momenta larger than the healing-length scale $k_\xi$ one observes the vortex-core scaling $\sim k^{-6}$. 
The above results reflect that in a vortex dominated flow, particles with low momenta are found far away from  vortex cores. 
In the case of pairing, the flow field far away from the cores is given by the field of a vortex pair, and the low-momentum scaling follows the pair-field scaling. Particles closer to the vortex cores pick up a higher momentum. Above $k_{\mathrm{pair}}$  the field will be dominated by the field of a single vortex. 
Note that \Fig{RandomVortex} shows the result of a numerical calculation in which we have sampled field configurations of random static vortices. 
Dynamical simulations of the classical field equation \Eq{GPE} will be shown in \Sect{Results}.

We close with recalling the picture Onsager developed in Ref.~\cite{Onsager1949a} of thermodynamic equilibrium states of a fixed number of vortices and antivortices in two dimensions.
He used the Hamiltonian of vortical flow in two dimensions \cite{Lin1941a},
\begin{eqnarray}  \label{eq:Hamiltonian}
H= -\frac{1}{2\pi} \sum_{i>j}^M \kappa_i \kappa_j \mathrm{ln}(|\mathbf{r}_{i} -\mathbf{r}_{j}|) \,,
\end{eqnarray}
to describe the dynamics of a system of $M$ vortices in a superfluid which hence behave like a Coulomb gas.
Here, the position of the $i$-th vortex is denoted as $\mathbf{r}_{i}=(x_i,y_i)$. 
Due to the fact that the $x$ and $y$ coordinates of each vortex are canonical conjugates, phase space is identical with configuration space of the vortex positions. 
Hence, for vortices moving in a  volume $V$ the total phase space is given by $V^M$. 
The Hamiltonian \eq{Hamiltonian} implies that low-energy configurations feature vortices of opposite sign close to each other, whereas high-energy configurations require vortices of equal sign to group. 
Due to these constraints, the number of configurations $\Omega(E)$ available for the system at a given energy $E$ decreases towards high and low energies, with a maximum at some intermediate $E=E_0$. 
According to Boltzmann, the entropy is $S(E)=k_B \mathrm{ln}(\Omega(E))$ and the inverse temperature $1/T=\del S/\del E$ is positive for $E<E_0$ and negative for $E>E_0$. 
It follows that positive-temperature states are characterized by vortex-antivortex pairing, while negative-temperature states feature vortices of the same circulation to cluster. 
At the point of maximum entropy $S(E_0)$ and infinite temperature, Onsager expected a state of uncorrelated vortices and antivortices. 
This understanding of the nonthermal fixed point as a quasi-equilibrium state of vortices corroborates our result that the IR scaling behavior at the fixed point corresponds to the appearance of topological excitations.

%
%
%======================================================================
\subsection{Vortex loops in $d=3$}
\label{subsec:VortexLoops}
%
%======================================================================
 \begin{figure}
 \includegraphics[width=0.45\textwidth]{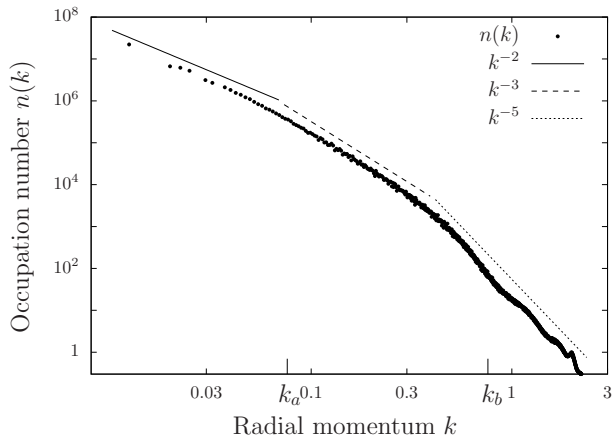}
 \caption{Radial momentum distribution of an elliptical vortex loop on a $N_s^{3}=1024^{3}$ grid. Note the double-logarithmic scale.
 The major and minor radius scales $k_{a}$, for $r_{a}=40$, and $k_{b}$, for $r_{b}=4$, respectively, are indicated.
 These scales separate pair scaling $n(k) \sim k^{-2}$ as for a near-circular vortex ring, 
 scaling $n(k) \sim k^{-3}$ for two anti-circulating vortex lines, and $n(k) \sim k^{-5}$, as for a vortex ring or of a pair of straight vortex lines, corresponding to the scaling exponent $\zeta_{Q}^{\mathrm{IR}}$ at the nonthermal fixed point.
 See \Subsect{VortexLoops} for more details.}
 \label{fig:Vortexellipse}
 \end{figure}
%======================================================================
%
%
We now consider the three-dimensional case of vortex lines and loops. 
A formulation similar to the Onsager point vortex model in \Subsect{IndependentVortices} is possible~\cite{Nemirovskii1998a,Nemirovskii2002a, Tsubota2008a}. 
See also Refs.~\cite{Chorin1991a, Berdichevsky1998a, Berdichevsky2002a, Nemirovskii2009a} for an extension including the dynamics of tangles of vortex lines. 
   
We write the classical kinetic energy spectrum of the velocity field in terms of the vorticity density 
\begin{eqnarray}  \label{eq:Vorticity}
\bm{\omega}(\mathbf{x}) = m\,\mathbf{\nabla}\times  \mathbf{v}(x) 
\end{eqnarray}
as
\begin{eqnarray}  \label{eq:EnergySpec3}
E_\mathrm{v}(\mathbf{k}) &=&   \frac{\langle| \bm{\omega}(\mathbf{k}) |^2\rangle}{2mk^2}  \, .
\end{eqnarray}
The vorticity is vanishing everywhere but on the vortex lines, i.e., assuming $M$ individual vortex loops, 
\begin{eqnarray}  
\bm{\omega}(\mathbf{x}) =  \sum_i^M \int_{0}^{L_{i}} \mathrm{d}\tau \, \mathbf{s}_i'(\tau) \delta(\mathbf{s}_i(\tau)-\mathbf{x}) \, .
\label{eq:vorticity}
\end{eqnarray}
In this expression, the vortex filaments are represented by the connected curves $\mathbf{s}_{i}(\tau)$, parametrized by the one-dimensional coordinate $\tau\in\{0,L_{i}\}$, $L_{i}$ being the arc length of filament $i$. 
The loops are closed, $\mathbf{s}_{i}(L_{i})=\mathbf{s}_{i}(0)$, possibly also accross the walls of the 3-dimensional volume in accordance with periodic boundary conditions.
$\mathbf{s}'_{i}(\tau)$ is the tangent vector along the filament at $\mathbf{s}_{i}(\tau)$, of unit length $|\mathbf{s}'_{i}(\tau)|=1$.
We parametrize the $i$-th vortex loop in terms of a single center coordinate and a relative curve,  $\mathbf{s}_i(\tau) = \mathbf{R}_i + \mathbf{r}_i(\tau)$ and write the vorticity as
\begin{eqnarray}  
\bm{\omega}(\mathbf{x}) 
=  \sum_i^M \int\, \mathrm{d}^{3}y\,\delta(\mathbf{y} - \mathbf{R}_i) \int_{0}^{L_{i}} \mathrm{d}\tau  \, 
\mathbf{r}_i'(\tau)\,
 \nonumber \\
\,  \times \delta(\mathbf{r}_i(\tau)  - \mathbf{x}+ \mathbf{y}) \, .
\end{eqnarray}
Hence, in Fourier space,
\begin{eqnarray}  
\bm{\omega}(\mathbf{k}) =  \sum_i^M  e^{i\mathbf{k}\mathbf{R}_i} \tilde{\bm{\omega}}_i(\mathbf{k})  \, ,
\end{eqnarray}
where $\tilde{\bm{\omega}}_i(\mathbf{k})$ is the Fourier transform of the vorticity of the $i$-th vortex loop, given by
\begin{eqnarray} 
\tilde{\bm{\omega}}_i(\mathbf{x}) =  \int_{0}^{L_{i}} \mathrm{d}\tau \, \mathbf{r}_i'(\tau) \delta(\mathbf{r}_i(\tau) - \mathbf{x}) \,.
\end{eqnarray}
The ensemble-averaged vorticity $\langle| \bm{\omega}(\mathbf{k}) |^2\rangle$ becomes
\begin{eqnarray} \label{eq:EnergySpecCorr3}
\langle| \bm{\omega}(\mathbf{k}) |^2\rangle = \sum_{i,j}^M  \langle e^{i\mathbf{k}(\mathbf{R}_i-\mathbf{R}_j)} \tilde{\bm{\omega}}_i(\mathbf{k}) \tilde{\bm{\omega}}_j(\mathbf{k}) \rangle \,.
\end{eqnarray}
Assuming that the shapes of the individual vortex loops are statistically independent of their position, it follows that $\langle e^{i\mathbf{k}(\mathbf{R}_i-\mathbf{R}_j)}  \tilde{\bm{\omega}}_i(\mathbf{k}) \tilde{\bm{\omega}}_j(\mathbf{k}) \rangle = \langle e^{i\mathbf{k}(\mathbf{R}_i-\mathbf{R}_j)} \rangle \, \langle \tilde{\bm{\omega}}_i(\mathbf{k}) \tilde{\bm{\omega}}_j(\mathbf{k}) \rangle$. 
If the loops are also uncorrelated among themselves, then $\langle \tilde{\bm{\omega}}_i(\mathbf{k}) \tilde{\bm{\omega}}_j(\mathbf{k}) \rangle =  \langle |\tilde{\bm{\omega}}_i(\mathbf{k})|^{2} \rangle \delta_{ij}$. Hence, for statistically identical loops,
\begin{eqnarray} \label{eq:EnergySpecCorrFinal}
\langle| \bm{\omega}(\mathbf{k}) |^2\rangle = {M} \langle  |\tilde{\bm{\omega}}(\mathbf{k})|^{2} \rangle \,,
\end{eqnarray}
which means that the vorticity spectrum scales in the same way as the average vorticity of a vortex loop centered at the origin. 
Finally, the scaling of the momentum spectrum $n(k)=2mk^{-2}E_\mathrm{v}(k)$ follows from that of the angle-averaged vorticity,  
\begin{eqnarray} \label{eq:SpectrumFromVorticity}
n(k) \sim k^{-4} \, \int \mathrm{d}\Omega_{k}\langle| \bm{\omega}(\mathbf{k}) |^2\rangle.
\end{eqnarray}
For the case of two straight parallel vortex lines of opposite circulation \Eq{EnergySpecCorrFinal} is evaluated in \Appendix{VortexLine}, for an ensemble of such paired lines in \Appendix{VortexLines}. Making use of the procedure developed there, the case of circular vortex rings of radius $r$ is discussed in \Appendix{VortexRing}. For the latter, the resulting angle-averaged momentum spectrum scales like $n(k)\sim k^{-2}$ for momenta $ k \ll k_r = 2\,\mathrm{sin}(\pi/2r)  $ and $n(k)\sim k^{-5}$ for momenta $k \gg k_r $. To include effects from squeezed vortex loops, we consider elliptical filaments. 
The ellipse is defined by a major radius $r_a$ and minor radius $r_b$. 
In \Fig{Vortexellipse}, the angle-averaged momentum spectrum of an elliptical vortex loop is shown. 
Three scaling regimes can be distinguished. For the lowest momenta, one has $n(k) \sim k^{-2}$, which equals the infrared scaling for a vortex ring. 
For momenta  $k_{a} \ll k \ll k_{b}$, one finds $n(k) \sim k^{-3}$, which coincides with the infrared scaling of two anti-circulating vortex lines (see \App{VortexLine}). 
For the ellipse, above $k_{b}$, the momentum distribution scales like $n(k) \sim k^{-5}$, which is the same as the high-momentum scaling of a vortex ring (see \App{VortexRing}) or of a pair of straight vortex lines (\App{VortexLines}).

%======================================================================
\subsection{Velocity distribution}
\label{subsec:SpectrumVelo}
%
%======================================================================
 \begin{figure}
 \includegraphics[width=0.45\textwidth]{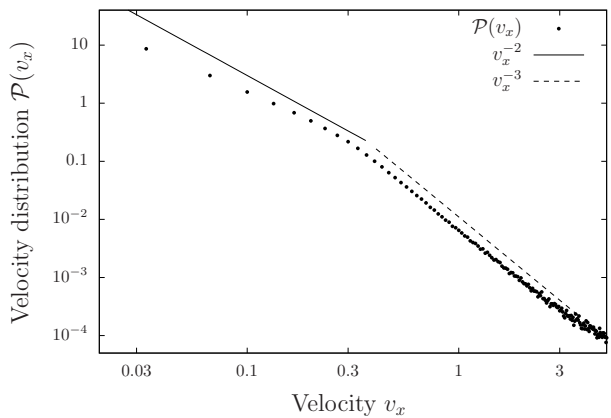}
 \caption{Velocity distribution \eq{Pvelx} as obtained from a random distribution of $M=80$ bound vortex-antivortex pairs, on a  $N_s^{2}=1024^{2}$ grid, as in \Fig{RandomVortex}. Note the double-logarithmic scale. At low momenta the velocity field follows the scaling of vortex-antivortex pairs ($\alpha=2$), whereas at higher momenta the distribution reflects the presence of vortices ($\alpha=1$). Note, that there is a deviation from $\alpha=2$ in the regime of low velocities. From simulations with higher resolution, we find the best agreement for $\alpha=1.8$. We remark, that in the limit $M \rightarrow \infty$ the velocity distribution is expected to be Gaussian \cite{Weiss1998a}.}
 \label{fig:VortexVelocity}
 \end{figure}
%======================================================================
%
%
The vortex model can provide insight into another observable accessible in our numerical simulations, the velocity distribution. See Refs. \cite{White2010a, Weiss1998a, Min1996a, Chavanis2001a} for discussions of the velocity distribution and Ref. \cite{Paoletti2008a} for recent experimental results. As discussed above, the velocity field scales, far away from any core, as $\sim r^{-\alpha}$, with $\alpha=1$ for a single vortex and $\alpha=2$ for a vortex-antivortex pair in two dimensions. The velocity probability distribution $\mathcal{P}(\mathbf{v})$ is calculated as $\mathcal{P}(\mathbf{v})=|\del\mathbf{x}/\del\mathbf{v}| \, \rho(\mathbf{x})$, with spatial vortex distribution function $\rho(\mathbf{x})$. For a uniform distribution $\rho(\mathbf{x}) = \mathrm{const.}$, it follows after angular averaging that
\begin{eqnarray}  \label{eq:Pvel}
\mathcal{P}(\mathbf{v}) &=& |\mathbf{v}|^{-2(\alpha+1)/\alpha} .
\end{eqnarray}
It is numerically convenient to calculate the probability density of a single component of the field, e.g.
\begin{eqnarray}  \label{eq:Pvelx}
\mathcal{P}(v_x) &=& \int \mathrm{d}v_y \, \mathcal{P}(\mathbf{v}) \simeq v_x^{1-2(\alpha+1)/\alpha} .
\end{eqnarray}
In \Fig{VortexVelocity}, the corresponding velocity distribution $\mathcal{P}(v_x)$ is shown. In accordance with the analytical predictions two scaling regimes are observed. At low momenta the velocity field follows the scaling of vortex-antivortex pairs, whereas at higher momenta the distribution reflects a random distribution of vortices.

%======================================================================
%====Bose gas approaching the nonthermal fixed point==============================================================
\section{Bose gas approaching the nonthermal fixed point}
\label{sec:Results}
In this section we return to the process of the formation of vortical excitations and of the Bose gas approaching the nonthermal fixed point characterized by the wave-turbulent scaling solutions discussed in \Sect{dyncritpoints} and Refs.~\cite{Scheppach:2009wu,Nowak:2010tm}. 
Thereby we first focus on the evolution of the single-particle momentum distributions and the emergence of nonthermal power laws. 
We relate these to structure formation in the form of vortical excitations. 
In order to identify in a clearer way the relevant processes in approaching the fixed point we compute the fluxes in momentum space. Remarkably, we find that the appearance of the particle scaling exponent $\zeta^{\mathrm{IR}}_{Q}$ in the IR and of the energy exponent $\zeta^{\mathrm{UV}}_{P}$ in the UV are compatible with predictions based on general arguments of irreversibility as well as energy and number conservation in a collision process~\cite{Zakharov1992a}.

%
%==================================================================
\subsection{Time evolution of the single-particle spectrum}
\label{subsec:SpectrumEvolution}
%

%======================================================================
 \begin{figure}[!t]
 \includegraphics[width=0.48\textwidth]{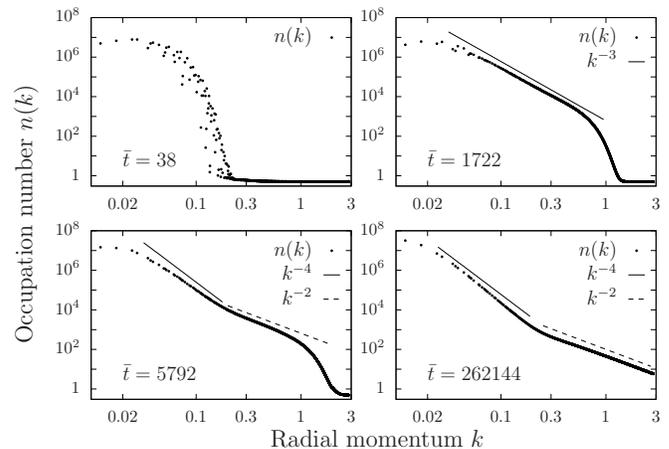}
 \caption{Single-particle mode occupation numbers as functions of the radial momentum $k$, for four different times of a run in $d=2$ dimensions. Parameters are: $\overline{g}=3\times 10^{-5}$, $N=4\times10^8$, $N_s=512$.
Note the double-logarithmic scale. 
An early development of a scaling $n(k)\sim k^{-3}$ is followed by a quasi-stationary period of bimodal scaling with $n(k)\sim k^{-4}$ in the IR, due to the presence of vortices, and $n(k)\sim k^{-2}$ in the UV, corresponding to weak wave turbulence or thermal equilibrium.}
\label{fig:ModeOccupationSequence2}
 \end{figure}
%======================================================================
%======================================================================
 \begin{figure}[!t]
 \includegraphics[width=0.48\textwidth]{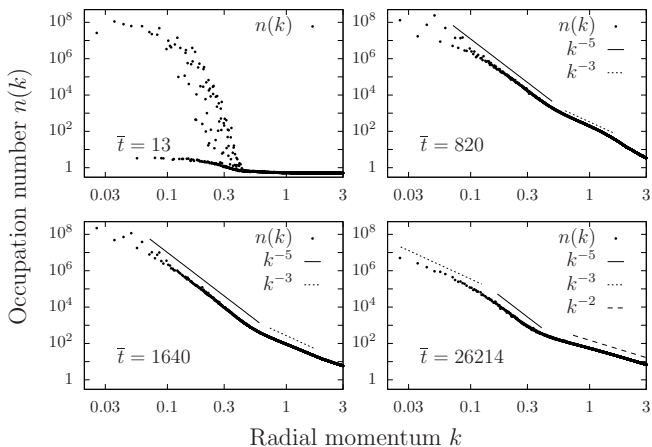}
 \caption{Single-particle mode occupation numbers as functions of the radial momentum $k$, for the four different times of a run in $d=3$ dimensions. Parameters are:  $\overline{g}=4\times 10^{-4}$, $N=8\times10^9$, $N_s=256$. 
Note the double-logarithmic scale. A period of bimodal scaling $n(k)\sim k^{-5}$ (vortex lines, IR) and $n(k)\sim k^{-3}$ (weak wave turbulence, UV) is followed by trimodal scaling, which also exhibits pairing, i.e., $n(k)\sim k^{-3}$ in the far IR, induced by a set of far separated small vortex rings, in addition to $n(k)\sim k^{-5}$ (IR), and thermal scaling $n(k)\sim k^{-2}$  in the UV.}
\label{fig:ModeOccupationSequence3}
 \end{figure}
%======================================================================
%
%
As discussed in \Subsect{SummaryVortexNumerics}, vortical excitations can be created in large numbers, within shock waves forming during the non-linear evolution of the coherent matter-wave field. 
Figs.~\fig{ModeOccupationSequence2} and \fig{ModeOccupationSequence3} show the angle- and ensemble-averaged radial momentum spectra, \Eq{spectrumdef}, for a Bose gas in a box with periodic boundary conditions, in $d=2$ and $d=3$ dimensions, respectively. 
Four snapshots are shown, taken at the dimensionless times $\bar t$ as indicated in each panel. 
See \Appendix{classfields} for details on the simulations and on lattice units.
The initial field configurations were prepared by macroscopically populating a few of the lowest momentum modes in the computation such that the resulting condensate density in configuration space varied between zero and some maximum value. 
In Figs.~\fig{ModeOccupationSequence2} and \fig{ModeOccupationSequence3}, the panels representing the earliest times show the system after a brief initial evolution during which momentum is rapidly transported from the few initially occupied modes near zero over a comparatively large range of wave numbers. 

Shortly after this, vortical excitations are created which immediately causes the spectrum in $d=2$ dimensions to exhibit a power-law behavior $2.85 \lesssim \zeta \lesssim 3.0$ within a range of momenta $k \in [0.04:0.4]$, see the upper right panel of Figs.~\fig{ModeOccupationSequence2}. 
Subsequently, the evolution slows down and a quasi-stationary period is entered. 
During an intermediate stage (bottom left panel of Fig.~\fig{ModeOccupationSequence2} and upper right panel of Fig.~\fig{ModeOccupationSequence3}) of the vortex-bearing phase two distinct power laws develop which are in excellent agreement with the analytical prediction in Eqs.~\eq{kappaUV} and \eq{kappaIR}. 
While in the ultraviolet the exponent $\zeta_{P}^{\mathrm{UV}} = d$ exhibits weak wave turbulence, \Eq{kappaUV}, in the infrared, the exponent confirms the field theory prediction $\zeta_{Q}^{\mathrm{IR}}=d+2$, cf.~\Eq{kappaIR}. 
More specifically, in two dimensions, at $\overline{t}=5792$, one observes scaling exponents $3.8\lesssim \zeta \lesssim 4.0$ within a range of momenta $k \in [0.02:0.2]$ and $2.0\lesssim \zeta \lesssim 2.3$ within a range of momenta $k \in [0.2:0.7]$, see the lower left panel of Figs.~\fig{ModeOccupationSequence2}. At $\overline{t}=262144$, $4.0 \lesssim \zeta \lesssim 4.2$ within a range of momenta $k \in [0.02:0.2]$, see the lower right panel of Figs.~\fig{ModeOccupationSequence2}. 
In $d=3$ dimensions, at $\overline{t}=820$, one observes $4.8\lesssim \zeta \lesssim 5.0$ within a range of momenta $k \in [0.08:0.4]$ and $3.0\lesssim \zeta \lesssim 3.1$ within a range of momenta $k \in [0.5:1.7]$, see the upper right panel of Figs.~\fig{ModeOccupationSequence3}. At $\overline{t}=1640$, $5.0 \lesssim \zeta \lesssim 5.1$ within a range of momenta $k \in [0.05:0.5]$, see the lower left panel of Figs.~\fig{ModeOccupationSequence3}.
The appearance of the bimodal power laws corroborates results for a relativistic $O(N)$ model reported in Refs.~\cite{Berges:2008wm,Berges:2010ez}. 

During the ensuing evolution, the weak-wave-turbulence scaling decays towards $\zeta=2$, reflecting a thermal UV tail.  
Note that in $d=2$, the weak-turbulence exponent $\zeta_{P}^{\mathrm{UV}}=2$ is identical to that in thermal equilibrium in the Rayleigh-Jeans regime, $n(k) \sim T/k^2$ \cite{Zakharov1992a}. 
In $d=3$ we observe, at late times, a change of the infrared scaling behavior from $\zeta=d+2=5$ to $\zeta=3$, pointing to the development of pairing correlations, cf.~Sects.~\subsect{CorrelatedVortices}, \subsect{VortexLoops}, and \Subsect{PairingSpectrum} below.
More specifically, at $\overline{t}=26214$, one observes a scaling exponent $2.9\lesssim \zeta \lesssim 3.0$ within a range of momenta $k \in [0.03:0.1]$, see the lower right panel of Figs.~\fig{ModeOccupationSequence3}.

At late times, after the last vortical excitations have disappeared, we observe the entire spectrum to become thermal, i.e., exhibit Rayleigh-Jeans scaling with $\zeta=2$ (not shown).
We emphasize that thermal scaling of the single-particle occupation number with $\zeta=2$ applies despite the fact that quasiparticles with a linear dispersion are expected to thermalize in the regime of wave numbers smaller than the inverse healing length $1/\xi \simeq 0.45$ ($d=2$) and $1/\xi \simeq 0.22$ ($d=2$). 
In the Bogoliubov approximation this is seen by taking into account the power-law dependence of the coefficients $u_{k}^{2}\sim v_{k}^{2}\sim k^{-1}$ which contribute to $n(k)\sim(u_{k}^{2}+ v_{k}^{2})T/k\sim T/k^{-2}$ \cite{Scheppach:2009wu}.

We are going to study more details of the time evolution of the system in a forthcoming publication and focus in the remainder of this paper on properties of the stationary scaling distributions. 
An important question in this context is, why the system selects the particular exponents $\zeta_{P}^{\mathrm{UV}} = d$ and $\zeta_{Q}^{\mathrm{IR}}=d+2$ from the set of four possible exponents given in Eqs.~\eq{kappaUV} and \eq{kappaIR}.
For this, the fluxes underlying the stationary but non-equilibrium distributions are relevant.

%==================================================================
\subsection{Fluxes}
\label{subsubsec:Fluxes}
%

%======================================================================
 \begin{figure}[t]
 \includegraphics[width=0.45\textwidth]{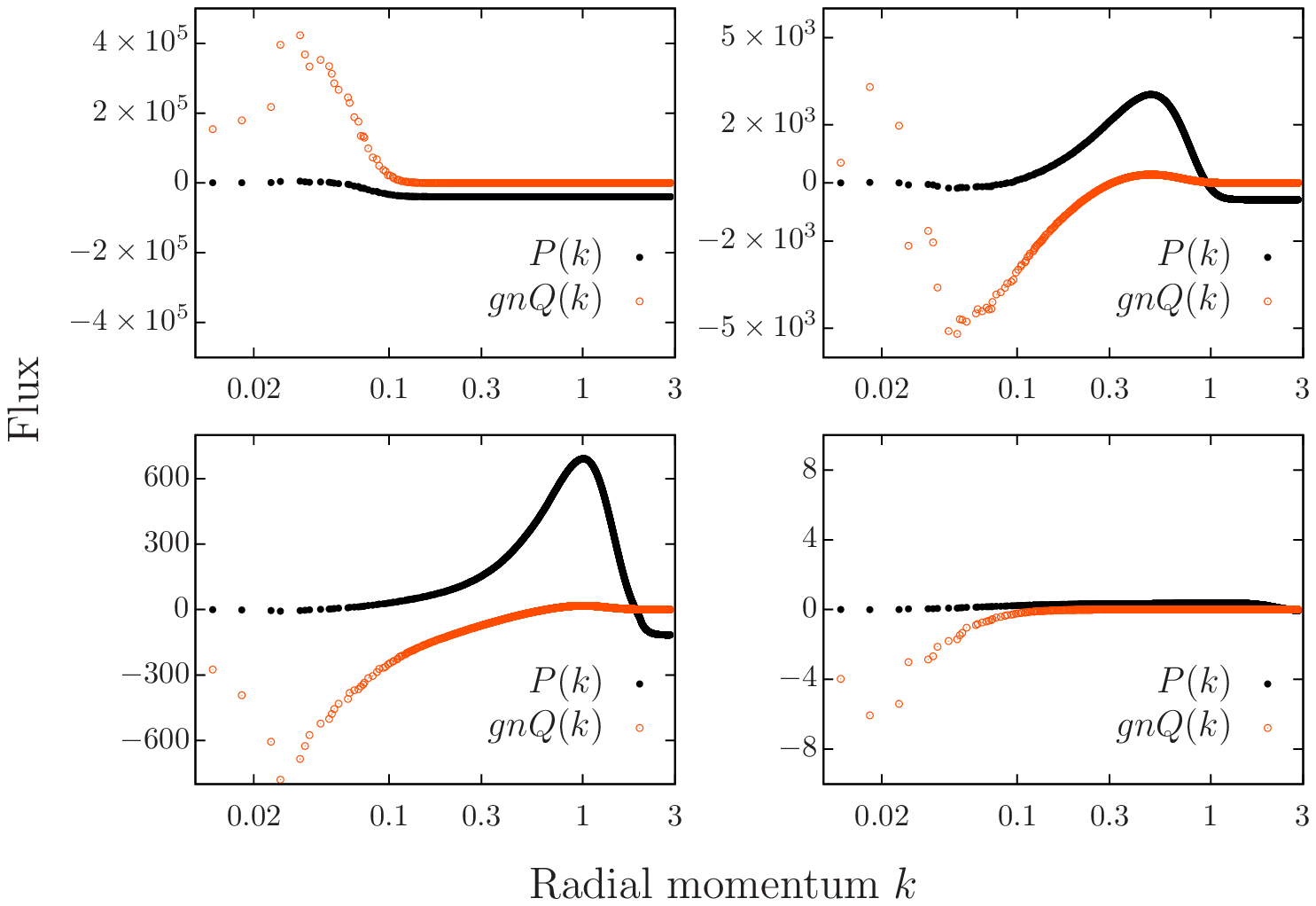}
 \vspace*{3ex}\\
 \includegraphics[width=0.45\textwidth]{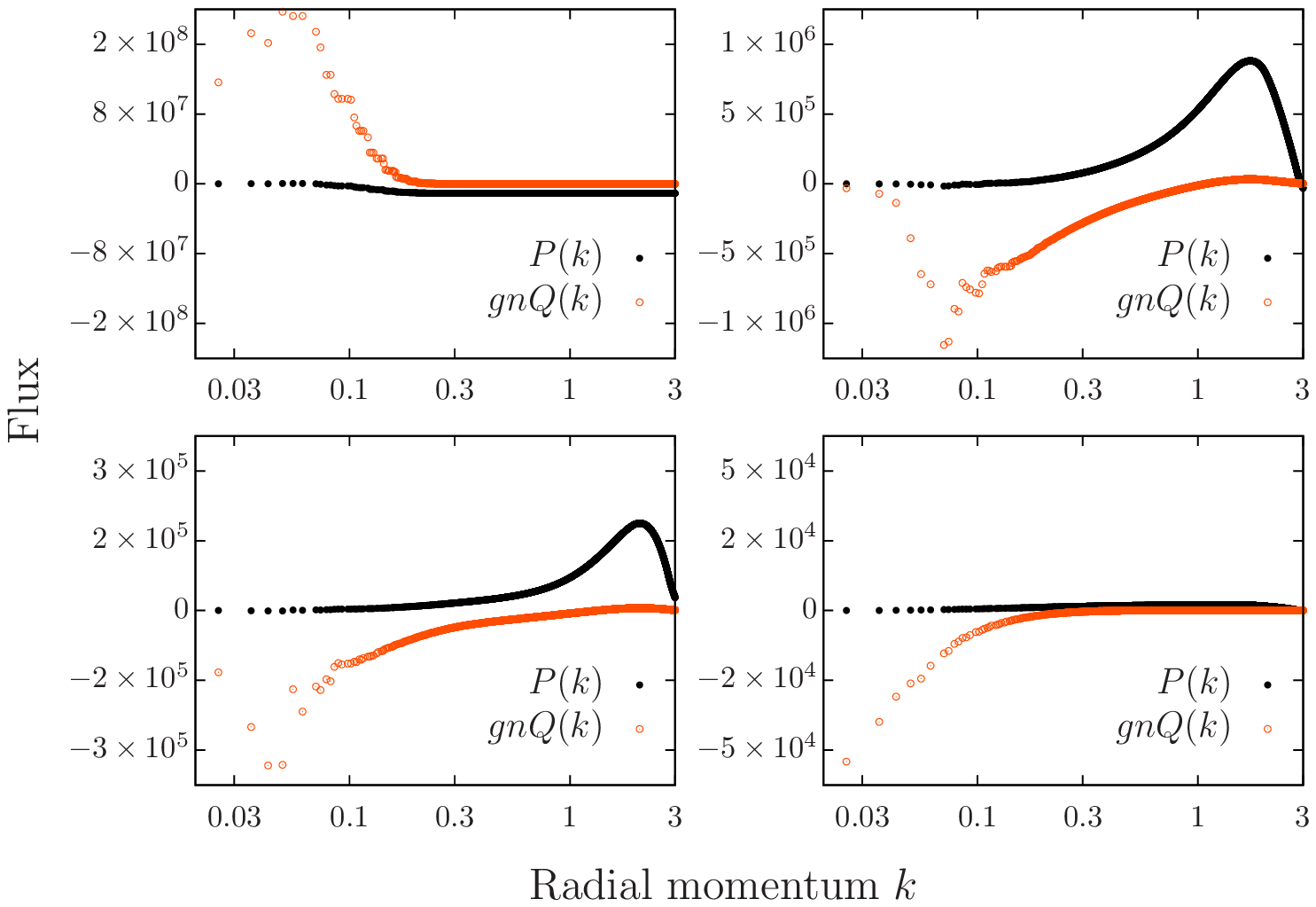}
 \caption{(Color online) Kinetic-energy and particle fluxes in $d=2$ (upper set) and $d=3$ (lower set), for the four different snapshots of Figs.~\fig{ModeOccupationSequence2} and \fig{ModeOccupationSequence3}, respectively. 
 Note the logarithmic $k$-axes. 
The appearance of the bimodal scaling coincides with a positive kinetic energy flux in the UV, and a negative particle flux in the IR.
Flux units: $[P]=[gnQ]=(4m^2a_s^{d+4})^{-1}$, cf.~\App{classfields}.
}
 \label{fig:PFlux2D3D}
 \end{figure}
%======================================================================
%
%
The timeline of distributions shown in Figs.~\fig{ModeOccupationSequence2} and \fig{ModeOccupationSequence3} suggests that the evolution of the gas involves a transport process from intermediate momenta around $0.05\ldots0.2$  both towards lower and higher wave numbers, building up a bimodal power-law distribution.
To describe the character of these transport processes we plot, in \Fig{PFlux2D3D} for $d=2$ (upper set of four panels) and 
$d=3$ (lower set), the radial particle and kinetic-energy flux distributions $Q_{k}$ and $P_{k}$, respectively. 
Note that the radial particle flux density $Q_{k}$ is multiplied by $gn$ to have the same units as the energy flux density $P(k)$. 
These flux densities are defined through the balance equations \eq{BalEqQ} and \eq{BalEqP}, respectively, with kinetic energy density $\varepsilon_k = n_k k^2/2m$. 
The initial stage is governed by an infrared particle transport towards larger $k$, also causing a flux $gnQ$ of interaction energy. 
At intermediate times, $Q(k)$ changes sign. 
This is accompanied by a positive kinetic energy transport in the UV, as observed in two-dimensional simulations in Ref.~\cite{Nazarenko2006a}. 

We emphasize that the negative particle flux in the IR and the positive kinetic-energy flux in the UV coincide with the appearance of the bimodal momentum distributions in Figs.~\fig{ModeOccupationSequence2} and \fig{ModeOccupationSequence3} (bottom left panels). 
Although the derivation of the IR exponents requires the full dynamical theory with non-perturbatively resummed self-energies, the signs of the fluxes correspond to the respective scaling exponents, i.e., $\zeta_{Q}^{\mathrm{IR}}$ in the IR, and $\zeta_{P}^{\mathrm{UV}}$ in the UV, cf.~\Subsect{wwt}.
Moreover, at late times, the kinetic-energy flux $P$ vanishes due to a thermalized UV momentum distribution, but $Q$ still reshuffles particles and therefore interaction energy, keeping the system out of equilibrium close to the nonthermal fixed point. 

We finally remark that, as discussed in detail in Ref.~\cite{Zakharov1992a}, one can show on the general grounds of the Boltzmann $H$-theorem that a necessary condition for a non-equilibrium stationary distribution in our systems is energy damping in the region of large $k$.
Moreover, energy and particle number conservation in the interaction of different momentum modes can be shown to imply the existence of at least one more sink, i.e., a region where the right-hand sides of \Eq{BalEqQ} effectively has an additional damping term $\sim \Gamma(k)n(k)$, with negative $\Gamma$.
In between these sinks, a source region supplies the input to the bidirectional flux pattern towards the UV and IR.
It was shown, moreover, in Ref.~\cite{Zakharov1992a}, within wave-kinetic theory, that generally, a positive, $k$-independent flux transports energy, $P>0$ while a negative flux transfers particles, $Q<0$. 
The only exception is the case of $d=2$ where the thermal and the weak-wave-turbulence exponent $\zeta_{P}^{\mathrm{UV}}$ can not be distinguished.
Remarkably, this pattern remains valid in our case, beyond the UV weak-wave-turbulence regime, in the IR region where the exponent emerges from a fixed point of the full dynamic equations.
As already pointed out in Ref.~\cite{Scheppach:2009wu}, however, the derivation of the IR exponent $\zeta_{Q}^{\mathrm{IR}}=d+2$  requires the existence of a sufficiently well defined quasi-particle dispersion relation, suggesting a treatment in terms of the Quantum Boltzmann equation with a momentum dependent scattering matrix element to be applicable.

From this point of view, the negative flux $Q$ and scaling in the IR and the positive flux $P$ and weak wave turbulence in the UV, as observed in the numerics, emerge as a necessary consequence of conservation laws and transport processes described by wave-kinetic transport equations.

%============================================================================================
%== Vortices, acoustic turbulence and the departure from the fixed point=====================
\section{Vortices, acoustic turbulence and the departure from the fixed point}
\label{sec:VorticesAnd}
In this last section we analyse the single-particle momentum spectra obtained in our numerical simulations in view of their interpretation in terms of vortical excitations and wave-turbulence as implied by the point- and line-vortex models introduced in \Sect{RandomVortexModel}.
For this we first decompose the flow pattern of the system which has closely approached the nonthermal fixed point, into transverse (incompressible) and longitudinal (compressible) contributions. 
In this way we can show that the IR scaling is dominated by the incompressible part while in the UV the particles mainly belong to the compressible as well as a quantum pressure components. 
We find a further scaling exponent $\zeta\simeq d+1$ for the subdominant compressible component in the IR which is interpreted as a signature of acoustic turbulence. 

Moreover, going beyond the results presented in \cite{Nowak:2010tm} we identify signs of pair formation in the final stage. 
The scaling caused by this goes beyond the predictions from dynamical quantum field theory summarized in \Sectionref{dyncritpoints}, and we interpret it as a signature for the system leaving the fixed point again for final thermalization.

%==================================================================
\subsection{Superfluid turbulence and statistics of vortices}
\label{subsec:SuperfluidTurbulence}
%

%======================================================================
 \begin{figure}[t]
 \includegraphics[width=0.45\textwidth]{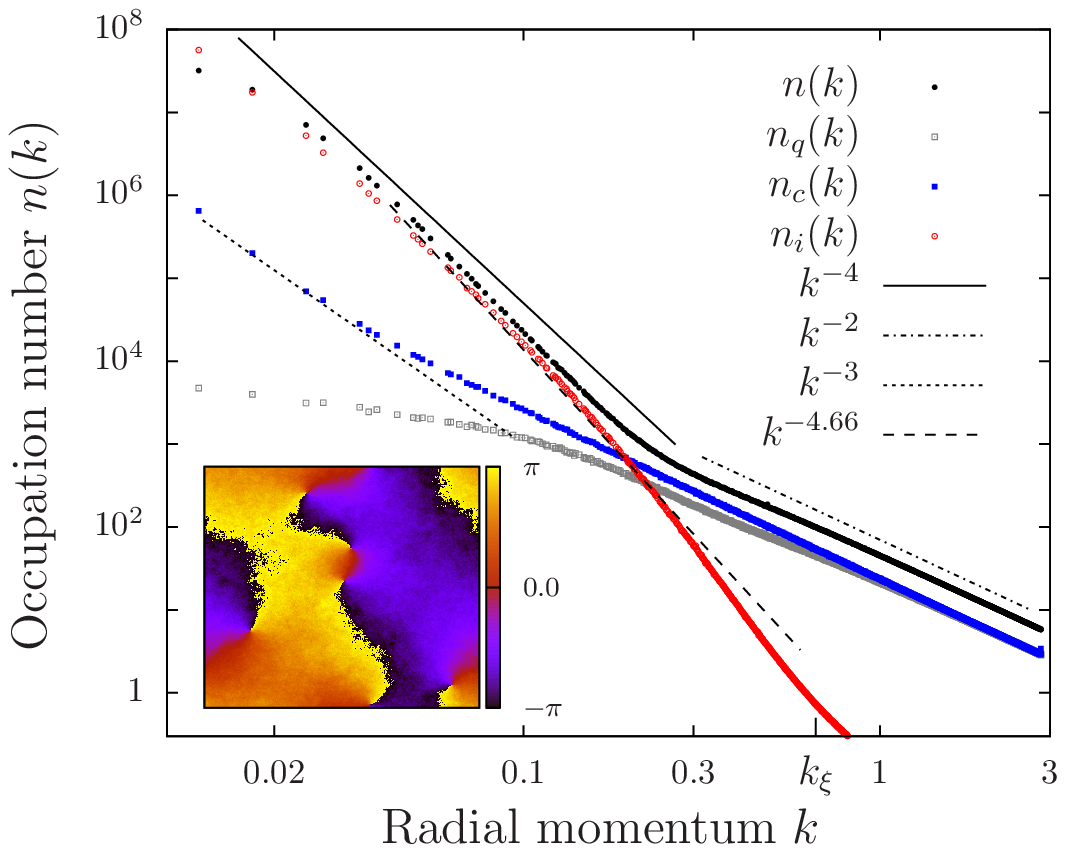}
 \caption{(Color online) Occupation numbers as functions of the radial momentum $k$, as defined in Eqs.~\eq{Ev}--\eq{decomposition}: 
 Total single-particle occupation number $n(k)$ (black dots), incompressible (solenoidal-flow) component $n_{\mathrm{i}}(k)$ (red circles), compressible (rotationless) component $n_{\mathrm{c}}(k)$ (blue filled squares), quantum-pressure component $n_{\mathrm{q}}(k)$ (grey open squares). 
 Parameters are the same as in \Fig{ModeOccupationSequence2}  (lower right panel), for the run in $d=2$ dimensions at the time $\overline{t}=262144$. Note the double-logarithmic scale. A scaling with $k^{-4.66}$ corresponds to a power-law exponent $5/3$ for the kinetic energy in $d=2$. See text for more details on the scaling exponents.
 Inset: Phase angle $\varphi(\vector{x},t)$ as in \Fig{2DPhase512}.}
 \label{fig:EnergySpectrum}
 \end{figure}
 \begin{figure}[t]
 \includegraphics[width=0.45\textwidth]{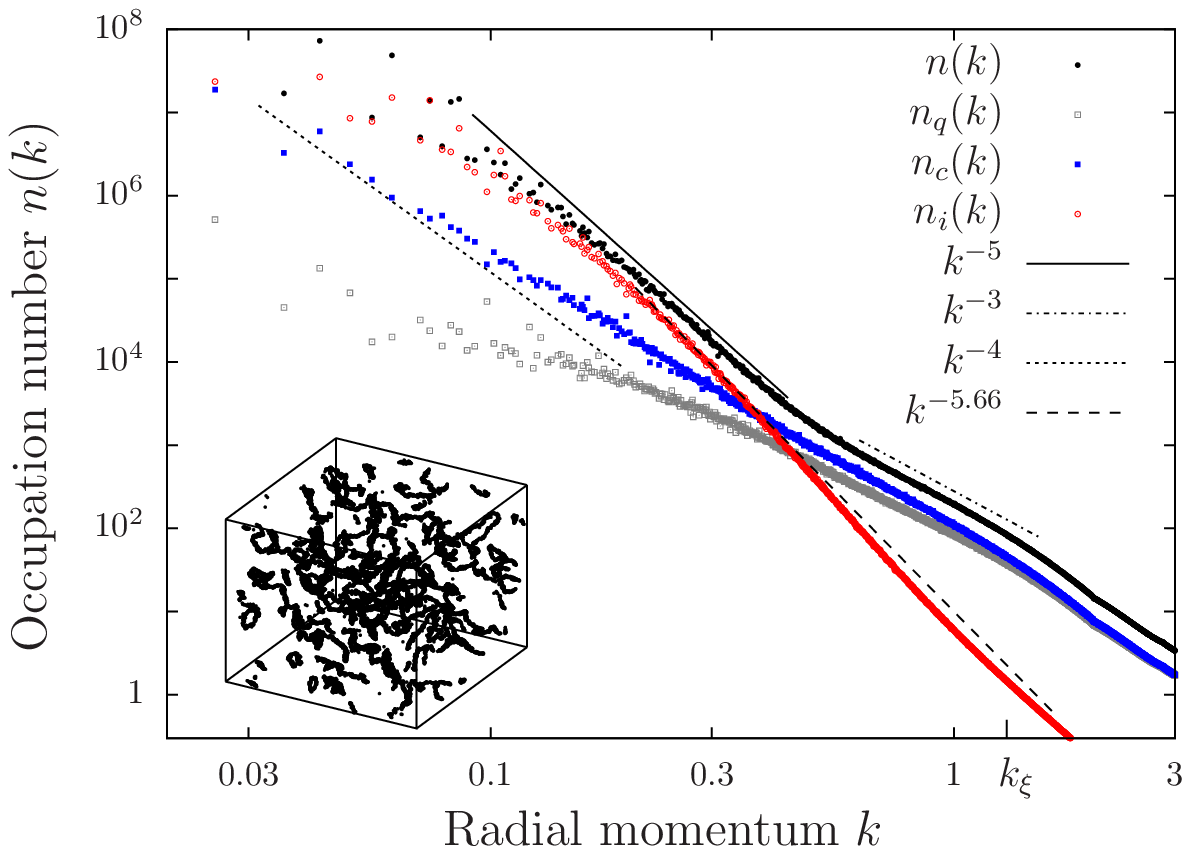}
 \caption{(Color online) Occupation numbers as functions of the radial momentum $k$, as defined in Eqs.~\eq{Ev}--\eq{decomposition} 
 (see caption of \Fig{EnergySpectrum} for more information).
 Parameters are the same as in \Fig{ModeOccupationSequence3}  (lower left panel), for the run in $d=3$ dimensions at the time $\overline{t}=1640$. Note the double-logarithmic scale.
A scaling with $k^{-5.66}$ corresponds to a power-law exponent $5/3$ for the kinetic energy in $d=3$. 
Inset: The black spots mark points around the vortex line cores where the density falls below 5\% of the mean density $n$.}
 \label{fig:ModeOccupationBoltzmann3}
 \end{figure}
%====================================================================== 
%

To exhibit vortical flow and define the decomposition we use the polar representation $\phi(\mathbf{x},t)=\sqrt{n(\mathbf{x},t)}\exp\{i\varphi(\mathbf{x},t)\}$ of the field in terms of the density $n(\mathbf{x},t)$ and a phase angle $\varphi(\mathbf{x},t)$. 
This allows to express the particle current $\mathbf{j}=i(\phi^*\nabla \phi - \phi \nabla \phi^*)/2=n\mathbf{v}$ in terms of the velocity field $\mathbf{v}=\nabla\varphi$.

With this, we decompose the kinetic-energy spectrum following Ref.~\cite{Nore1997a,Nore1997b}, splitting the total kinetic energy $E_{\mathrm{kin}}= \int \mathrm{d}^dx \, \langle |\nabla \phi(\mathbf{x},t)|^2\rangle/(2m)$ as $E_{\mathrm{kin}} = E_{\mathrm{v}} + E_\mathrm{q}$ into a `classical' part 
\begin{equation}
\label{eq:Ev}
E_\mathrm{v}= \frac{1}{2m}\int \mathrm{d}^dx \, \langle |\sqrt{n}\mathbf{v}|^2 \rangle 
\end{equation}
and a `quantum-pressure' component 
\begin{equation}
E_\mathrm{q}=\frac{1}{2m}\int \mathrm{d}^dx \, \langle |\nabla \sqrt{n}|^2 \rangle \,.
\end{equation}
The radial energy spectra for these fractions involve the Fourier transform of the generalised velocities $\vector{w}_{\mathrm{v}}=\sqrt{n}\vector{v}$ and $\vector{w}_{\mathrm{q}}=\nabla\sqrt{n}$,
\begin{eqnarray}
 E_{\delta}(k)= \frac{1}{2m} \int k^{d-1} \mathrm{d}\Omega_d \, \langle |\mathbf{w}_{\delta}(\mathbf{k})|^2 \rangle,\quad \delta=\mathrm{v},\mathrm{q}.
\end{eqnarray}
Following Ref.~\cite{Nore1997a,Nore1997b}, the velocity $\mathbf{w}_{\mathrm{v}}$, which due to the multiplication of $\mathbf{v}$ with the density $n$ becomes regularized and can be Fourier transformed, is furthermore decomposed into `incompressible' (divergence free) and `compressible' (solenoidal) parts, $\mathbf{w}_{\mathrm{v}}=\mathbf{w}_{\mathrm{i}}+\mathbf{w}_{\mathrm{c}}$, with $\nabla \cdot \mathbf{w}_\mathrm{i}=0$, $\nabla \times \mathbf{w}_\mathrm{c}=0$, to distinguish vortical superfluid and rotationless motion of the fluid. For comparison of the kinetic-energy spectrum with the single-particle spectra $n(k)$, we determine occupation numbers corresponding to the different energy fractions as 
\begin{equation} 
\label{eq:decomposition}
n_\mathrm{\delta}(k) =  k^{-d-1}E_{\mathrm{\delta}}(k)\,,\,\, \delta \in \{\mathrm{i}, \mathrm{c}, \mathrm{q}\}. 
\end{equation}
The resulting spectra $n_\mathrm{i}(k)$, $n_\mathrm{c}(k)$, and $n_\mathrm{q}(k)$ add up to $n_\mathrm{s}(k) = n_\mathrm{i}(k) + n_\mathrm{c}(k) + n_\mathrm{q}(k)$, which agrees with the single-particle spectrum up to small corrections, see \Appendix{hydro}.

In Figs.~\fig{EnergySpectrum} and \fig{ModeOccupationBoltzmann3}, we depict the momentum distributions of the occupation numbers $n_\mathrm{i}(k)$, $n_\mathrm{c}(k)$, and $n_\mathrm{q}(k)$, together with the previously shown total single-particle spectrum $n(k)$, each at a late time when the system is close to the fixed point, i.e., shows the predicted scaling both in the IR and the UV.
Red circles denote $n_{\mathrm{i}}$, filled blue squares $n_{\mathrm{c}}$, and open grey squares $n_{\mathrm{q}}$. 
Qualitatively, the results are similar for $d=2$ and $d=3$. 

In the range of large wavenumbers, the spectrum is dominated by the compressible and quantum-pressure components. 
The scaling of these excitations exhibits the weak-wave-turbulence exponent $\zeta_{P}^{\mathrm{UV}} = d$. 
For smaller wave numbers the scaling changes to $n(k)\sim k^{-d-2}$.
The decomposition into the various components now shows that this switching to a different scaling in the IR is clearly due to the take-over of a different character of the excitations with a modified flow pattern accounted for by the incompressible (solenoidal) component of $\mathbf{w}_{\mathrm{v}}$.
The fact that in this regime contributions from vortical flow $n_{\mathrm{i}}$ dominate is in accordance with our interpretation of the strong IR scaling by a model of an ensemble of vortical excitations.
As we have discussed in the previous section the analytically predicted infrared power laws $n(k)\sim k^{-d-2}$ are consistent with a finite density of independent vortices and antivortices ($d=2$) or vortex lines ($d=3$).

Moreover, we find that the compressible component in the IR represents the second strongest contribution to the flow and develops a nonthermal scaling as $\sim k^{-d-1}$. 
We will discuss this result in more detail in \Subsect{AcousticTurbulence} below.

For momenta larger than about $k\simeq0.2$ for $d=2$ and $k\simeq0.4$ for $d=3$ the compressible and quantum-pressure components dominate the momentum distributions.
In the regime of intermediate momenta, above the scale $k_l \sim 2\pi/l$ with $l$ the mean inter-vortex spacing, where the incompressible flow and the rest are of roughly equal strength one finds a scaling of approximately $n_\mathrm{i}(k)\sim k^{-d-1-5/3}$, corresponding to $E_{i}(k)\sim k^{-5/3}$ for both the two- and the three-dimensional case.
Scaling with this exponent, which is the same as in classical Kolmogorov turbulence in an incompressible fluid, has been found in simulations of superfluid turbulence in $d=2, 3$ before \cite{Nore1997a,Nore1997b,Araki2002a,Kobayashi2005a,Kobayashi2005b,Horng2007a,Horng2009a,Tsubota2008a,Numasato2010a}.
In some of these cases, turbulent flow appeared in the time evolution of a system starting from a Taylor-Green configuration of vortex line tangles. 
We emphasize that our configuration after the creation of vortical excitations resembles more a state of the kind usually termed chaotic turbulence.
It is unclear whether chaotic turbulence can be related to classical Kolmogorov turbulence as this does not bear near-classical bundles of equal-orientation vortex lines. 
Nonetheless we observe such $5/3$ scaling in the range of intermediate momenta where the interplay between the incompressible and compressible components is most pronounced.

%
%== Results: Quasistationary stage ================================================================
\subsection{Acoustic turbulence}
\label{subsec:AcousticTurbulence}
%
%
%======================================================================
 \begin{figure}[t]
 \includegraphics[width=0.45\textwidth]{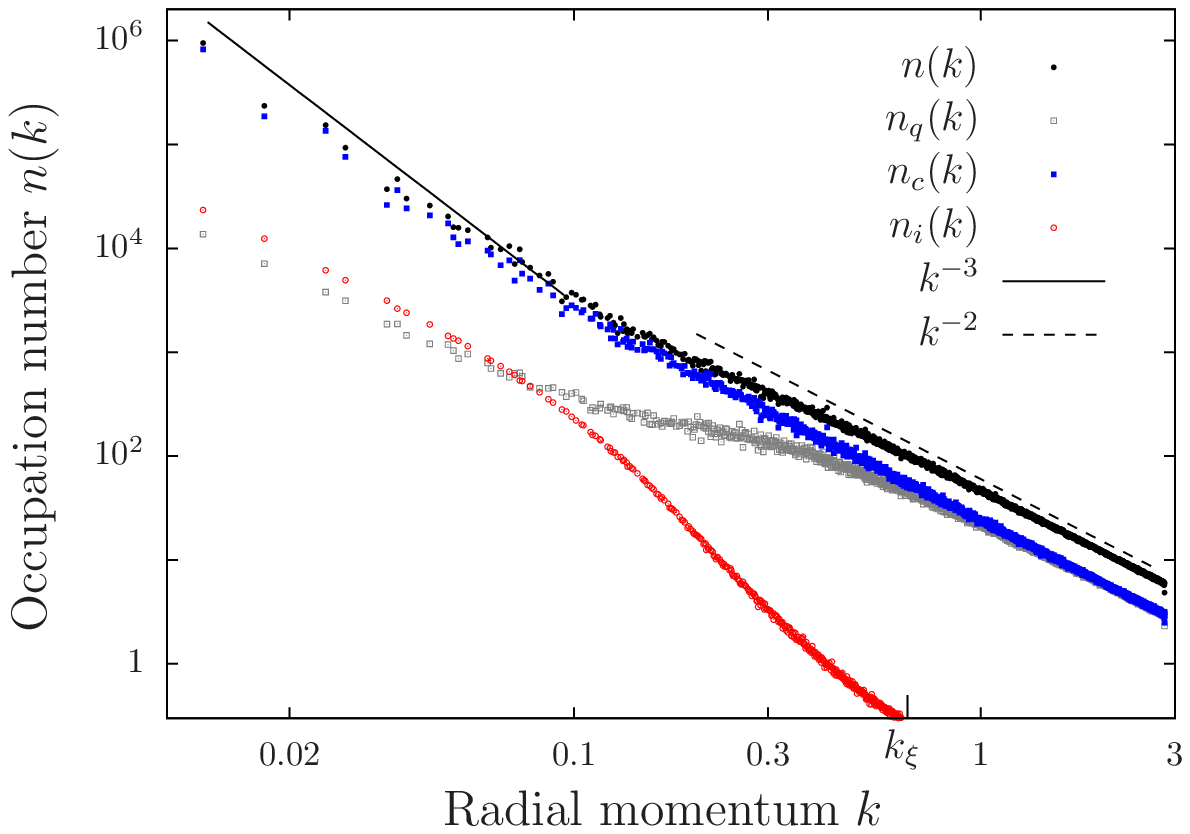}
 \vspace*{0.5cm}\ \\
 \includegraphics[width=0.45\textwidth]{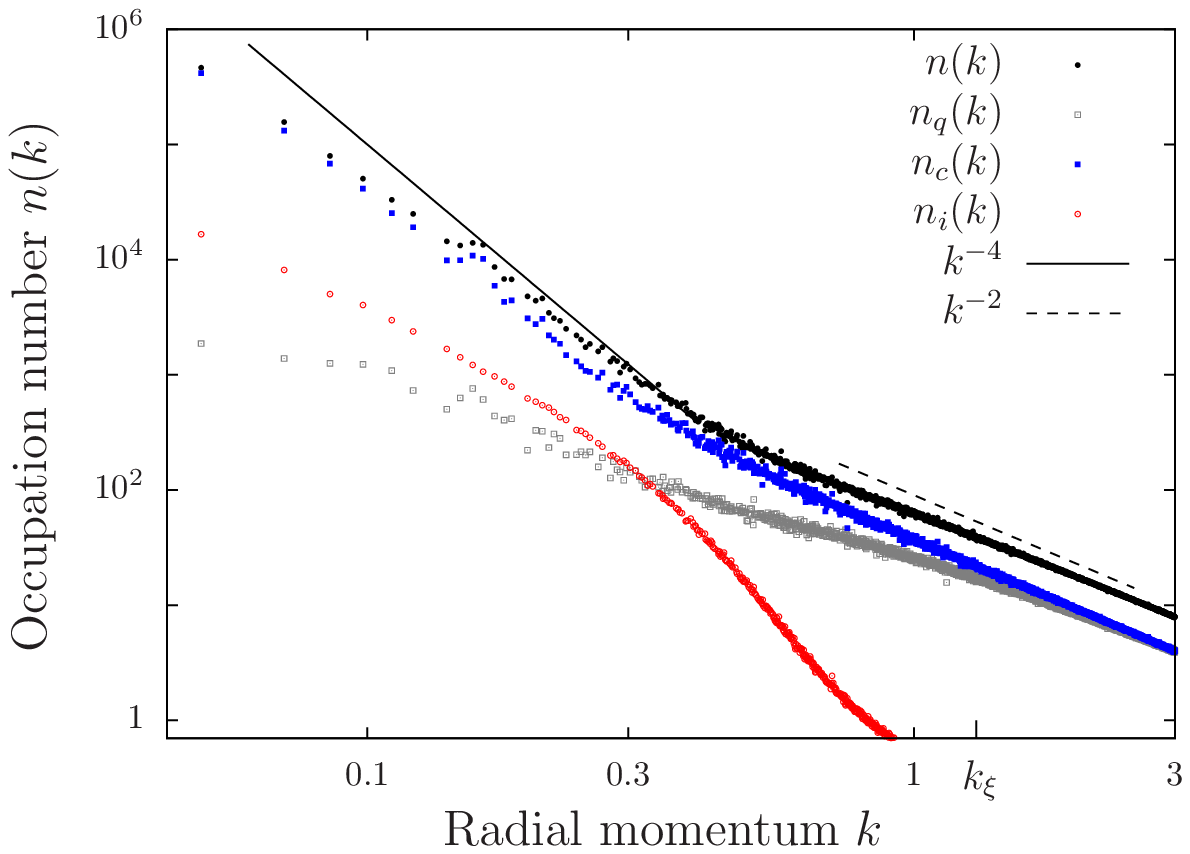}
 \caption{(Color online) Late-stage acoustic turbulence: 
 Occupation numbers defined in Eqs.~\eq{Ev}--\eq{decomposition} are shown as functions of the radial momentum $k$, from an average over $\sim10$ single runs in $d=2$ dimensions (upper panel) and $d=3$ (lower panel), at times $\overline{t} = \mathcal{O}(10^6)$ ($d=2$) and $\overline{t} = \mathcal{O}(10^4) $ ($d=3$), triggered on the decay of the last vortical excitation.
See caption of \Fig{EnergySpectrum} for more information.
  Parameters are:  $d=2$: $\overline{g}=3\times 10^{-5}$ , $N=4\times 10^8$, $N_s=512$; $d=3$: $\overline{g}=4\times 10^{-4}$, $N=10^9$, $N_s=128$. Note the double-logarithmic scale. The figure shows that, shortly after the last vortex ring has disappeared (incompressible component breaks down), compressible excitations exhibiting acoustic turbulence scaling $\sim k^{-(d+1)}$ remain present.}
 \label{fig:Acoustic}
 \end{figure}
%======================================================================
%
%
The IR scaling $\sim k^{-d-1}$ of the compressible component (blue filled squares) in Figs.~\fig{EnergySpectrum}, \fig{ModeOccupationBoltzmann3}, and also \Fig{PairingSpectrum} below suggests an interpretation in terms of acoustic turbulence \cite{Zakharov1992a,Kagan1992a,Kagan1994a,Kozik2009a} and corroborates the numerical findings reported in Ref.~\cite{Khlebnikov2002a}. 
The scaling is persistent until late times, see, e.g., \Fig{PairingSpectrum}, but decays after the vortices have disappeared. To check that the nonthermal scaling of the incompressible component is not an artefact of the decomposition of the vortical flow into compressible and incompressible parts, we show the momentum spectrum for an average over selected runs in $d=3$, see \Fig{Acoustic} (lower panel), where the snapshot is taken shortly after the last vortex ring has disappeared. One can observe that acoustic turbulence can survive for a limited period, $\Delta\overline{t}=\mathcal{O}(100)$, beyond the time when vortical excitations are negligible. In the case of $d=2$, the effect is weaker but still present, see \Fig{Acoustic} (upper panel), where we show the averaged spectrum shortly after the last vortex-antivortex annihilation. Our interpretation of this finding is that acoustic turbulence coexists with the dominant vortical flow. It is driven by vortex motion and especially vortex annihilation processes, which are known to produce compressible excitations. In $d=2$, signatures of acoustic turbulence are less pronounced, which we attribute to reduced vortex dynamics or, equivalently, weaker driving of compressible excitations. 
%

%== Results: Quasistationary stage ================================================================
\subsection{Vortex velocities}
\label{subsubsec:Vortexvelocities}
%
%
%
%======================================================================
 \begin{figure}[t]
 \includegraphics[width=0.45\textwidth]{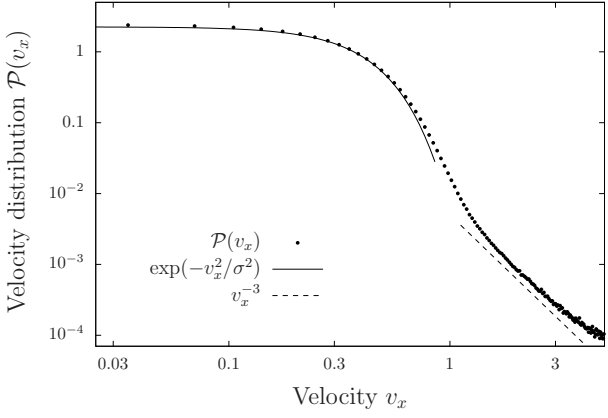}
 \caption{Velocity field probability distribution. Parameters are the same as in \Fig{ModeOccupationSequence3}, for the run in $d=3$ dimensions, at the time $\overline{t}=1640$. Note the double-logarithmic scale. The black line is a Gaussian fit to the data which shows that in the low-velocity regime, the distribution is dominated by Gaussian fluctuations. A high-velocity scaling with $v_{x}^{-3}$ reflects the presence of uncorrelated vortices, cf.~\Eq{Pvelx}.}
 \label{fig:VelocityDistribution3}
 \end{figure}
 \begin{figure}[t]
 \includegraphics[width=0.45\textwidth]{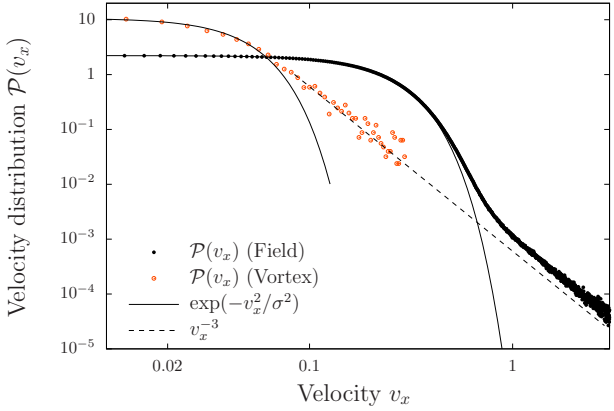}
 \caption{(Color online) Comparison of the vortex velocity distribution with the velocity-field probability distribution in $d=2$ dimensions at $\overline{t}=32768$. At this time an average of $86$ vortices are present in the system. Parameters: $\bar g= 3\times 10^{-5},$ $N=16\times 10^8$, $N_s=1024$. Note the double-logarithmic scale.
 A high-velocity scaling with $v_{x}^{-3}$ reflects the presence of uncorrelated vortices, cf.~\Eq{Pvelx}.
 Gaussian fluctuations are suppressed in the distribution of velocities of the individual vortices, measured through the motion tracking of the vortex cores.}
 \label{fig:VelocityDistribution2}
 \end{figure}
%======================================================================
%
We furthermore investigated the velocity-field probability distribution in the turbulent regime to compare our data with the expectations summarized in \Subsect{SpectrumVelo}. 
In $d=3$ dimensions, see \Fig{VelocityDistribution3}, the results corroborate theoretical and experimental results reported in Refs.~\cite{White2010a, Paoletti2008a, Daniels2003a, Proment2010a}. 
The low-momentum distribution is characterized by Gaussian statistics, whereas the UV regime clearly shows scaling $\mathcal{P}(v_{x})\sim v_{x}^{-3}$ predicted by the model of randomly distributed, uncorrelated vortex lines, see \Eq{Pvelx}. 
Note that even at late times, when vortex rings shrink in size, no signs of vortex pairing, $\mathcal{P}(v_{x})\sim v_{x}^{-2}$, can be seen in the velocity distribution. 
This is due to dominating Gaussian fluctuations in the low momentum region. 
In $d=2$ dimensions, see \Fig{VelocityDistribution2}, we compare the velocity-field distribution with the velocity distribution of the positions of the individual vortices. Similar to the three-dimensional case the IR part of the velocity-field distribution is characterized by Gaussian statistics, whereas the UV regime shows single-vortex scaling~\cite{White2010a}. 
The vortex velocity distribution is obtained from the statistical analysis of a vortex-tracking algorithm  which is designed to detect regions of low density accompanied by a winding number equal to one. The method is analogous to the experimental set-up of Ref.~\cite{Paoletti2008a}, where particles are trapped inside vortex cores to study vortex velocities in $d=3$. In agreement with the experiment, \Fig{VelocityDistribution2} shows that Gaussian random motion is suppressed for vortices, but $\sim v^{-3}$ scaling clearly persists. In this way, quantum turbulence can be distinguished from classical turbulence, where a continuous distribution of vorticity favors a Gaussian velocity distribution.
%

%======================================================================
\subsection{Pairing and departure from the fixed point}
\label{subsec:PairingSpectrum}
%
%
%======================================================================
 \begin{figure}
 \includegraphics[width=0.45\textwidth]{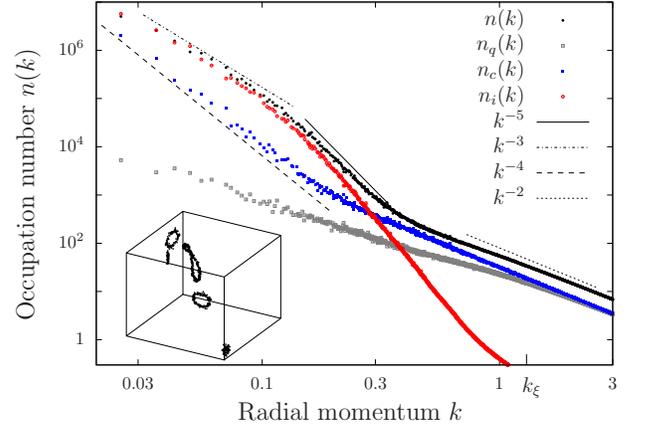}
 \caption{(Color online) Pairing effects: Occupation numbers are shown as functions of the radial momentum $k$. Parameters are the same as in \Fig{ModeOccupationSequence3}, for the run in $d=3$ dimensions, at the time $\overline{t}=26214$ (lower right panel). Note the double-logarithmic scale. 
 The scaling $\sim k^{-3}$ in the far IR reflects pair correlations present in far-separated small vortex loops.
(See caption of \Fig{EnergySpectrum} for more information).}
 \label{fig:PairingSpectrum}
 \end{figure}
 \begin{figure}[t]
 \includegraphics[width=0.45\textwidth]{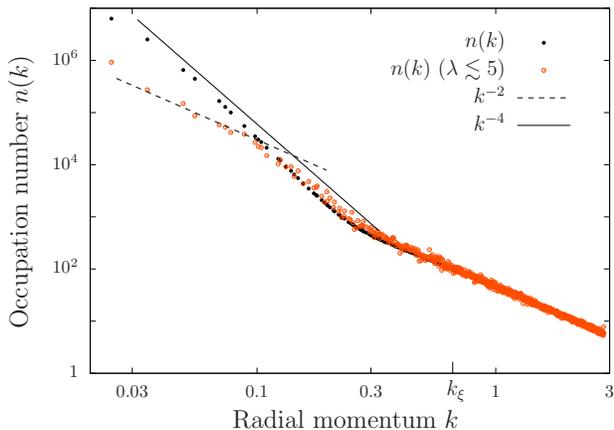}
 \caption{(Color online) Pairing effects: Occupation numbers are shown as functions of the radial momentum $k$ in $d=2$ dimensions, at the time $\overline{t}=262144$: Occupation number obtained from averaging over all runs (black dots); Occupation number obtained from averaging over selected runs featuring tightly bound vortex antivortex pairs with maximum pair correlation length $\lambda \lesssim 5$ (red circles). Parameters: $\bar g = 3\times 10^{-5}$, $N=10^8$, $N_s=256$. Note the double-logarithmic scale. 
 A scaling with $k^{-2}$ reflects the presence of correlated vortex-antivortex pairs in $d=2$.}
 \label{fig:PairingSpectrum2}
 \end{figure}
%======================================================================
%
%
In \Fig{PairingSpectrum} we show the decomposition of the single-particle spectrum into the previously defined incompressible, compressible, and quantum-pressure components, see \Eq{decomposition}, for the 3-dimensional system at $\bar t = 26214$, as in the lower right panel of \Fig{ModeOccupationSequence3}. At this time, only a small density of vortex rings has survived, which decays under the influence of the noise field, cf. also Refs. \cite{Leadbeater2001a, Berloff2007a}. During the period when only a few small vortex rings remain one can observe a decrease of the infrared scaling exponents of the occupation number and its incompressible part to the value $\zeta=3$. Since vortical excitations are still dominating the spectrum, we can interpret this observation in terms of the statistical point vortex model introduced in \Sect{RandomVortexModel}, as vortex-antivortex pair correlations, see \Subsect{VortexLoops}. This is consistent with the snapshots of individual runs showing small vortex ellipses (see inset of \Fig{PairingSpectrum}). The scaling transition can be identified with the scale $k_b=2\pi/2r_b$, with estimated minor radius $r_b\simeq 15$. 

\begin{figure}[t]
 \includegraphics[width=0.45\textwidth]{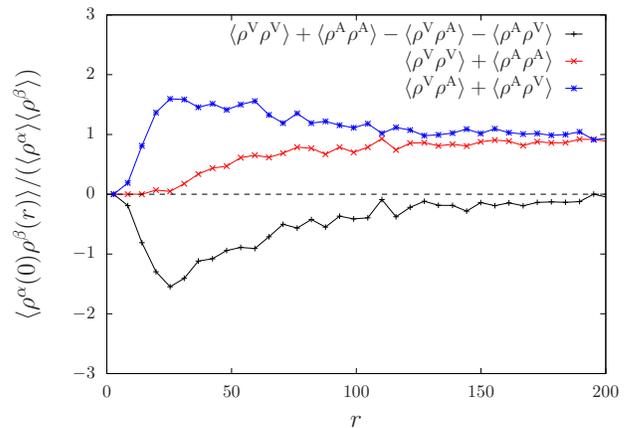}
 \caption{(Color online) Normalized pairing (density-density) correlations of vortices and antivortices in $d=2$. Parameters as in \Fig{ModeOccupationSequence2}, at the time $\overline{t}=131072$.
  Red crosses: Pair correlations of vortices and antivortices within the same circulation.
  Blue stars: Pair correlations between vortices and antivortices. The rise above $1$ indicates an effective ``binding''.
  Black crosses:  Resulting correlation function $C$ introduced in \Eq{CF}.
}
 \label{fig:PairCorr2D}
 \end{figure}

\Fig{PairingSpectrum2} shows the average occupation number spectrum at $\overline{t}=262144$ in $d=2$ dimensions, as in the lower right panel of \Fig{ModeOccupationSequence2}. In the average over all generated runs pairing effects can hardly be observed. 
However, for selected runs with small maximum pair correlation length $\lambda \lesssim 5$, as introduced in \Subsect{CorrelatedVortices}, the pair scaling appears. 
We find that in the selected runs ($\sim 0.2 \%$ of total number of runs), at the chosen point of time one or at most a few vortex-antivortex pairs with pairing length smaller than the distance between pairs are present. 
This constitutes the final period of the evolution, shortly before the last pair has disappeared through mutual annihilation and the system fully thermalizes.
The generic configuration in $d=2$ dimensions during the preceding time interval of critical slowing down close to the nonthermal fixed point is characterized by randomly positioned vortices and antivortices, with pairing correlations nevertheless present.
Pairing is seen in the density-density correlation function between vortices and antivortices shown in \Fig{PairCorr2D}, from which we read off a largest scale $\lambda_{\mathrm{max}}\simeq100\ldots150$, cf.~\Subsect{CorrelatedVortices}.
This implies pair scaling below $k\simeq0.02\ldots0.03$, not visible in the few (black) full circles in \Fig{PairingSpectrum2}.
%
%======================================================================
%======================================================================
%
We observe that, in $d=2$, scattering between the vortices of equal and opposite circulation can lead to the increase of the pairing length between correlated vortices and antivortices.
This is in contrast to the situation for $d=3$ where the scattering between vortex rings rarely leads to an increase in the size of the rings.
It is energetically advantageous for the rings to shrink in size, and thus, at a particular late time of the evolution the presence of separated small rings is generic and ''pairing'' is seen in the spectrum.

In summary, our analysis gives a picture of the necessary conditions and the character of the nonthermal fixed point. 
It also shows that pairing effects are a first signature for the system leaving this fixed point again for the final move to thermal equilibrium.

%
%
%== Summary ================================================
\section{Summary}
\label{subsec:Summary}
We have studied in detail superfluid turbulence in two- and three-dimensional dilute Bose gases by means of simulations in the classical-wave limit of the underlying quantum field theory. 
A focus is set on the identification and characterization of stationary scaling solutions, in particular for single-particle momentum distributions. 
Characteristic exponents $\zeta$ of $n(k)$ corroborate the analytical predictions made in Ref.~\cite{Scheppach:2009wu}. 
Our findings suggest that local field expectation values and short- to intermediate-range coherence, including topological excitations, are at the basis of the infrared power laws predicted within a full nonperturbative dynamical field theory \cite{Berges:2008wm,Berges:2008sr,Berges:2008mr,Scheppach:2009wu,Berges:2010ez,Gasenzer:2011by} employing two-particle-irreducible effective action techniques.

We have shown that the stationary scaling is maintained by the presence of particle (IR) and energy (UV) fluxes, originating at intermediate scales and directed towards the low- and high-frequency limits, respectively.
The respective fluxes were found to be consistent with the particular scaling exponents of the momentum distribution in the respective regimes.
Our main result is the identification of the nonthermal fixed-point with the appearance of topological excitations in the system.
We could successfully employ statistical models of point vortices in two dimensions and vortex rings in three dimensions to interpret the IR scaling exponent of the spectra. 
Moreover, we observed in our simulations, during the final thermalization stage, decay of the turbulence scaling into a scaling derived from the assumption of vortex-antivortex pairing. 
To characterize further the bimodal power spectra developed close to the nonthermal fixed point we have analysed the decomposition of the overall flow pattern into compressible and incompressible components.
We found that the IR power spectra of underlying compressible excitations suggest an understanding in terms of acoustic turbulence on top of the vorticity-bearing quasicondensate. 
Finally, to compare with the results of recent experiments and analytical predictions, the velocity-field probability distribution as well as the velocity statistics of individual vortices has been studied. 
This observable is of great interest, as it can be used to experimentally distinguish between classical and quantum turbulence.

The connection of QT phenomena with ab-initio dynamical field theoretic methods points a way to unified analytical studies of turbulence.
Moreover, it provides hints of how the proposed nonthermal fixed points  in relativistic systems \cite{Berges:2008wm,Berges:2008sr, Berges:2008mr,Berges:2010ez, Carrington:2010sz} are realized in nature. 
A manifestation of the approach of a nonthermal fixed-point in terms of quasi-solitary pattern formation and charge separation in a relativistic scalar model has recently been reported in Ref.~\cite{Gasenzer:2011by}.
 
Experimental studies of universal phenomena in nonequilibrium dynamics of ultracold atoms have great potential since universal effects
do not depend significantly on initial conditions and details of the system. 
Following this idea, our numerical protocol was chosen such that an experimental verification of our findings is within reach of present-day cold-atom experiments.  
The study of turbulence in ultracold gases may have great impact on many other fields of physics. 
Prominent examples are strongly correlated nuclear matter produced in heavy-ion collisions and early-universe cosmology.

%==============================================================================

\acknowledgments
T.~G. and D.~S. thank J. Berges and C. Scheppach for collaboration on related work. 
The authors thank J.~Berges, N.~Berloff, E.~Bodenschatz, M.~J.~Davis, G.~Falkovich, S.-C.~Gou, H.~Horner, R.~Kerr, G.~Krstulovic, M.~K.~Oberthaler, J.~M. Pawlowski, B.~Shivamoggi, B.~Svistunov, and M.~Tsubota for useful discussions. 
They acknowledge support by the Deutsche Forschungsgemeinschaft (GA 677/7,8), by the University of Heidelberg (FRONTIER, Excellence Initiative, Center for Quantum Dynamics), by the Helmholtz Association (HA216/EMMI), and by the University of Leipzig (Grawp-Cluster).
The authors furthermore thank KITP, Santa Barbara, for its hospitality.
This research was supported in part by the National Science Foundation under Grant No.~PHY05-51164.

%== Appendix ================================================
\appendix
\label{Appendix}
%
%
%==============================================================================
\section{Vortical excitations in a superfluid}
%==============================================================================
\subsection{Single vortex in $d=2$ dimensions}
\label{app:SingleVortex}
A singly quantized vortex with circulation $\kappa= \pm 1 $ in two dimensions is described, in polar coordinates, by the solution $\phi(r,\varphi)=f(r)\exp\{i\kappa\varphi\}$ of the Gross-Pitaevskii equation \eq{GPE}, where $f(r)$ is real and approaches the square root of the bulk density $n_{\mathrm{bulk}}$ for large distances $r$ from the vortex core where $f(0)=0$. 
A vortex is a stationary solution of \Eq{GPE}, evolving as $\phi(r,\varphi,t)=\phi(r,\varphi,0)\exp\{-i\mu t\}$ with  $\mu=gn_{\mathrm{bulk}}^{1/2}$. 
The velocity field of the vortex is 
\begin{equation}  \label{eq:vrphi}
\tilde{\mathbf{v}}(r,\varphi) = \frac{\kappa}{mr} \mathbf{e}_{\varphi} \,.
\end{equation}
The Fourier transform of this field can be conveniently evaluated by the help of the first-order Bessel function, $\int \mathrm{d}\varphi \, \mathrm{cos}(\varphi) \exp\{i\alpha\mathrm{cos}\varphi\} = 2 \pi i J_1(\alpha)$, as
\begin{align}
\tilde{\mathbf{v}}(k,\varphi_{k}) 
&=  \frac{\kappa}{m} \int \mathrm{d}r \, d\varphi \begin{pmatrix} -\mathrm{sin}(\varphi) \\ \mathrm{cos}(\varphi) \end{pmatrix} e^{ikr\mathrm{cos}(\varphi-\varphi_k)}
\nonumber\\
&=\frac{\kappa}{m} \frac{2\pi i}{k} \mathbf{e}_{\varphi_{k}}.
\end{align}
%

%==============================================================================
\subsection{Straight vortex line in $d=3$}
\label{app:VortexLine}
\begin{figure}
 \includegraphics[width=0.45\textwidth]{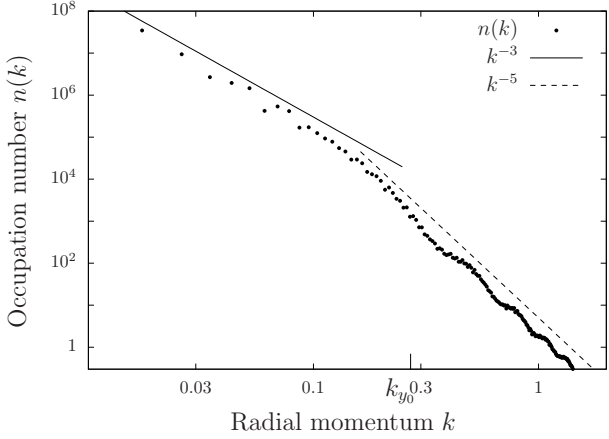}
 \caption{Radial momentum distribution of two straight vortex lines of circulation $\kappa_1=-\kappa_2=1$ on a $N_s^3=1024^3$ lattice. The distance scale $k_{y_0}$ is indicated. Note the double-logarithmic scale.}
 \label{fig:Straightvortexline}
 \end{figure}

Based on \Eq{vorticity}, we calculate the vorticity of two straight vortex lines in three dimensions, with circulation $\kappa_i$, $i=1,2$, placed at positions $(0,-y_0)$ and $(0,y_0)$ in the $x$-$y$ plane and relate it to the momentum spectrum, see \Eq{SpectrumFromVorticity}. The lines are parametrized by 
\begin{eqnarray} 
\mathbf{s}_i(\mathbf{x}) = \left( 0,  (-1)^i y_0,  \kappa_i \tau \right)
\end{eqnarray}
with $\tau \in [-\infty,\infty]$. 
In Fourier space, the vorticity density, defined in \Eq{Vorticity}, reads
\begin{eqnarray} 
\bm{\omega}(\mathbf{k}) =   \mathbf{e}_z\delta(k_z)  ( \kappa_1 e^{-ik_y y_0} + \kappa_2 e^{ik_y y_0}) \,.
\end{eqnarray}
The vorticity spectrum follows from \Eq{vorticity},
\begin{eqnarray} 
|\bm{\omega}(\mathbf{k})|^2 = 2\delta^2(k_z) \left[ 1 + \kappa_1\kappa_2 \mathrm{cos}(2 k_y y_0) \right]\,.
\end{eqnarray}
Hence, for co-rotating vortex lines ($\kappa_1=\kappa_2$)
\begin{eqnarray} \label{eq:Lines1}
|\bm{\omega}(\mathbf{k})|^2 = 4\delta^2(k_z) \mathrm{cos}^2(k_y y_0) \,,
\end{eqnarray}
while for counter-rotating vortex lines ($\kappa_1=-\kappa_2$) one obtains
\begin{eqnarray} \label{eq:Lines2}
|\bm{\omega}(\mathbf{k})|^2 = 4\delta^2(k_z) \mathrm{sin}^2(k_y y_0) \,.
\end{eqnarray}
In order to be able to take the angle average over $\delta^2$ we regularize the delta distribution, giving it a finite width $\Delta=1/L_z$.
The angle average of \Eq{Lines1} then reads
\begin{eqnarray}
|\bm{\omega}(k)|^2 
&= &
\int \mathrm{d}\varphi_k \mathrm{d}u \, \delta_{\Delta}(ku)^{2}
\nonumber \\ 
&&\quad \times
4\cos^2(k y_0 \sqrt{1-u^{2}} \mathrm{cos}\varphi_k )  
\nonumber\\
&=& 8\pi^{2} L_z k^{-1}[1+J_{0}(2k y_0)],
\end{eqnarray}
where $L_z$ is the length of the system in $z$-direction and $u=\mathrm{cos}(\theta)$.
One obtains the scaling behaviour
\begin{equation}
|\bm{\omega}(k)|^2 = 16\pi^{2}L_z k^{-1} +\mathcal{O}(k),
\end{equation}
which is the scaling of a single line. With \Eq{SpectrumFromVorticity}, we conclude $n(k)\sim k^{-5}$.
Following the same reasoning, the angle average of \Eq{Lines2} is given as
\begin{eqnarray} 
|\bm{\omega}(k)|^2 
&=& 8\pi^{2} L_z k^{-1}[1-J_{0}(2k y_0)]\,.
\end{eqnarray}
For $ ky_{0}\ll 1$
\begin{eqnarray}
|\bm{\omega}(k)|^2 
&=& 8\pi^{2}L_z y_{0}^{2}k +\mathcal{O}(k^{3})\,,
\end{eqnarray}
which gives $n(k)\sim k^{-3}$. For $ ky_{0}\gg 1$
\begin{eqnarray} 
|\bm{\omega}(k)|^2 
&=& 8\pi^{2} L_z k^{-1}
\end{eqnarray}
which gives $n(k)\sim k^{-5}$.
These scalings are confirmed in \Fig{Straightvortexline}.
%
%
%======================================================================
%======================================================================
%
%
%======================================================================
 \begin{figure}
 \includegraphics[width=0.45\textwidth]{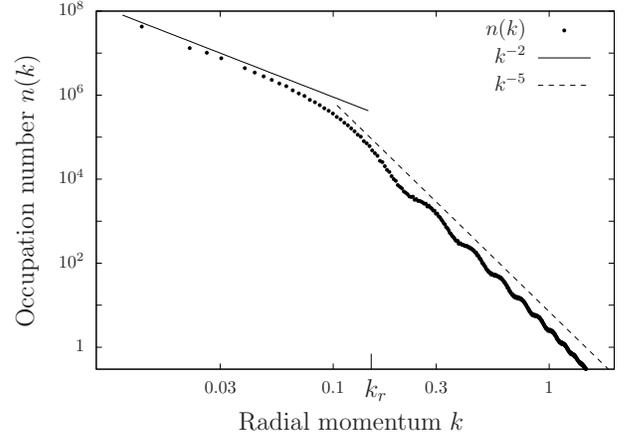}
 \caption{Radial momentum distribution of a vortex ring of radius $r$ on a $N_s^3=1024^3$ lattice. The radius scale $k_{r}$ is indicated. Note the double-logarithmic scale.}
 \label{fig:Vortexring}
 \end{figure}
%======================================================================
%

%======================================================================
\subsection{Ensembles of parallel vortex lines in $d=3$}
\label{app:VortexLines}
In Sects.~\subsect{IndPairs} and \subsect{CorrelatedVortices}, we have discussed the scaling of the momentum distribution for ensembles of independent and pair correlated vortices in two dimensions, respectively.
In this appendix we generalize the scaling derived above for straight vortex lines to the case of many such lines in $d=3$.
We derive an expression for the momentum spectrum of a system of $M$ straight vortex lines oriented along the $z$ direction. 
The average vorticity squared is given by \Eq{EnergySpecCorr3} where $\mathbf{R}_i$ are vectors in the two-dimensional $x$-$y$ plane pointing to the $i$-th vortex line with circulation $\kappa_i = {\pm 1}$.
The vorticity of the single line is
\begin{eqnarray} 
\tilde{\bm{\omega}}_i(\mathbf{k}) 
&=& \kappa_i \int \mathrm{d}^3x \, e^{i\mathbf{k}\mathbf{x}} \int \mathrm{d}\tau \, \mathbf{e}_z   \delta(\kappa_i\tau\mathbf{e}_z - \mathbf{x}) \, , \nonumber\\ 
&=&  \kappa_i \delta(k_z) \mathbf{e}_z  \, .
\end{eqnarray}
Hence, \Eq{EnergySpecCorr3} has the same form as \Eq{Vortexdensity} for $d=2$,
\begin{eqnarray} \label{eq:EnergySpecCorr3Line}
\langle| \bm{\omega}(\mathbf{k}) |^2\rangle 
=  \delta^2(k_z) \sum_{i,j}^M  \langle \kappa_i \kappa_j e^{i\mathbf{k}(\mathbf{R}_i-\mathbf{R}_j)}  \rangle \,,
\end{eqnarray}
up to a $\delta^2(k_z)$ term, which arises from the infinite extent of the vortex lines in $z$-direction. 
As in \App{VortexLine}, we regularize the delta distributions, and averaging \eq{EnergySpecCorr3Line} over solid angles yields
\begin{eqnarray} \label{eq:EnergySpecCorr3LineFinal}
\langle| \bm{\omega}(k) |^2\rangle &\sim& \frac{L_z}{k} \sum_{i,j}^M  \langle \kappa_i \kappa_j J_0(k|\mathbf{R}_i-\mathbf{R}_j|) \rangle \,,
\end{eqnarray}
which is analogous to \Eq{VE} in two dimensions. 
As the position of the $i$-th vortex line is determined by $\mathbf{R}_i$, statistical averaging of \eq{EnergySpecCorr3LineFinal} can be done in the same way as in $d=2$.

%==============================================================================
\subsection{Vortex ring in $d=3$}
\label{app:VortexRing}
A vortex ring with radius $r$ lying in the $x$-$y$ plane can be parametrized as
\begin{eqnarray} 
\mathbf{s}(\mathbf{x}) = \left( r\cos\varphi ,  r\sin\varphi , 0  \right)
\end{eqnarray}
with $\varphi \, \in \, [0,2\pi]$. In Fourier space, similar steps as above lead to the angle-averaged vorticity density 
\begin{eqnarray} 
|\bm{\omega}(k)|^2 =  \int \mathrm{d}\theta_k \, |\mathrm{J}_1(kr\sin\theta_k)|^2 \,.
\end{eqnarray}
Evaluating the integral numerically shows that the vorticity scales like $\sim k^2$ for $k r \ll 1$ and  $\sim k^{-1}$ for $k r \gg 1$, corresponding to $n(k)\sim k^{-2}$ and $n(k)\sim k^{-5}$. 
This is shown in \Fig{Vortexring}. 
Both scalings are also found in the momentum spectrum of a vortex ellipse, discussed in \Subsect{VortexLoops}.
%

%
%== Semiclassical field simulations ================================================================
\section{Semiclassical field simulations}
\label{app:classfields}
In this paper, a dilute superfluid gas of Bosons of mass $m$ is considered, with contact interactions quantified by a coupling constant $g$.
Its dynamics is described, in the classical wave limit, by the Gross-Pitaevskii equation (GPE) \eq{GPE}.
We use units where $\hbar=1$. 
We consider the gas to be contained in a box of size $L^d$, $d=2,\,3$, with periodic boundary conditions. 
In $d=3$, the coupling constant $g$ is related to the $s$-wave scattering length $a$ by $g=g_{\mathrm{3D}}=4\pi a/m$.
In $d=2$, one has $g=g_{\mathrm{2D}}=-(4\pi/m)[\ln(\mu ma_{\mathrm{2D}}^{2}/4)]^{-1}$ where $\mu$ is the chemical potential and $a_{\mathrm{2D}}$ is a scattering length in two dimensions which, for a gas of hard-spheres of radius $a$ is given by $a_{\mathrm{2D}}=ae^{\gamma}$, with the Euler-Mascheroni constant $\gamma\simeq0.577$  \cite{Schick1971a,Fisher1988a,Lee2002a}. 
For a two-dimensional gas created by trapping a three-dimensional one tightly in one dimension, with harmonic-oscillator length $l_{z}$, the effective 2D scattering length is given by $a_{\mathrm{2D}}=4l_{z}(\pi/B)^{1/2}\exp\{-\sqrt{\pi}l_{z}/a_{\mathrm{3D}}\}$, where $B\simeq0.915$ \cite{Petrov2000a}.
Diluteness implies that $a\ll l$, the interparticle spacing $l=n^{-1/d}$ being determined by the mean density $n=N/L^d$. 

The initial values for the real and imaginary parts of the field $\phi(\mathbf{k},0)$ are randomly chosen from a Gaussian  distribution with width $1/2$, centered around $\sqrt{n(\mathbf{k},0)}\exp\{i\varphi(\mathbf{k},0)\}$, where $n(\mathbf{k},t)=\langle \phi^\dag(\mathbf{k},t)\phi(\mathbf{k},t)\rangle$ is the occupation number at time $t$ and $\varphi(\mathbf{k},0)$ is a random phase angle. 
Correlation functions including $n(\mathbf{k},t)$ are obtained by averaging over many trajectories. 
To induce transport from small to large wave numbers, only a few modes near $\mathbf{k}=0$ are chosen to be macroscopically occupied at the initial time, $n(\mathbf{k},0)\gg1$. 
Such an initial state can be prepared, e.g., by Bragg scattering of photons from a Bose-Einstein condensate. 

Our numerical simulations are performed on space-time lattices with side lengths $L=N_s a_s$, with lattice spacing $a_s$ and in total $N_s^d$ gridpoints. 
We use periodic boundary conditions  and $N_s\in\{256, 512\}$ for $d=2$ and $N_s\in\{128, 256, 512\}$ for $d=3$. 
\Eq{GPE} is written in terms of the dimensionless variables  $\overline{g}=2mga^{2-d}_s$, $\overline{t}= t/(2ma^2_s)$ and $\overline{\psi}_n(t)=\psi_n \sqrt{a^d_s} \mathrm{exp}(2i\overline{t})$. 
All quantities in the figures are either dimensionless or shown in lattice units, with the length unit given by $a_{s}$.
The dimensionless lattice momenta are $k=[\sum_{i=1}^d 4\mathrm{sin}^2 (k_i/2)]^{1/2}, \mathbf{k}=2\pi\mathbf{n}/N_s, \, \mathbf{n}=(n_1,...,n_d),\,  n_i= -N_s/2,...,N_s/2$.

In order to relate our simulations to a typical situation in experiment, we give parameters for Rb-87. 
We estimate the total energy of the gas in equilibrium to be given by the interaction energy which,
by equipartition is related to the temperature, 
$2E_{\mathrm{int}} = L^d gn^2=N_s^d a^d_s gn^2 =N_s^d k_BT$.
At the transition to degeneracy the thermal de Broglie wave length $\lambda_\mathrm{dB} = \sqrt{{2\pi\hbar^2}/{mk_{B}T}}$ is of the order of the interparticle spacing and hence $k_BT_{\mathrm{deg}} \sim 2\pi\hbar^2n^{2/d}/m$.
Inserting this into the expression for the interaction energy allows to express the lattice spacing in terms of the density and the scattering length as
\begin{align}
a_s &= \frac{1}{2n}\sqrt{\ln4-\ln(\mu ma_{\mathrm{2D}})}\quad (d=2),
\nonumber\\
a_s &= ({n^{4/3}a} )^{-1/3}\quad (d=3).
\end{align}
Inserting parameters for a typical Rb-87 experiment in $d=3$, viz.,
$a = 5$\,nm, $m= 1.4\times 10^{-25}$kg,  
$n=10^{20}$\,m$^{-3}$ we obtain $a_s=1$\,$\mu$m. 
Hence, time scale as $t/\overline{t}= 2ma_s^2/\hbar^2= 3\times 10^{-3}$\,s. 
Our lattice time typically runs until $\overline{t} \simeq 1000$, which therefore represents a realistic observation timescale in experiments.  
Typical parameter choices in our simulations are $N=10^9$, $N_s=128$, $\overline{g}=4\times 10^{-4}$ for $d=3$ and
$N=4\times 10^8$, $N_s=512$, $\overline{g}=3\times 10^{-5}$ for $d=2$.

%== Semiclassical field simulations ================================================================
\section{Decomposition of $\vector{w}$}
\label{app:hydro}
%
%
%======================================================================
 \begin{figure}
 \includegraphics[width=0.48\textwidth]{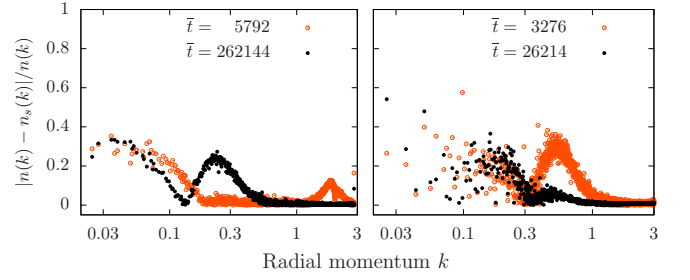}
 \caption{(Color online) Normalized occupation number difference $|n(k)-n_\mathrm{s}(k)|/n(k)$ as a function of the radial momentum $k$ at different times, for $d=2$ (left panel) and $d=3$ (right panel). 
Parameters are the same as in \Fig{ModeOccupationSequence2}, for the run in $d=2$ dimensions and in \Fig{ModeOccupationSequence3}, for $d=3$. Note the  logarithmic scale. }
 \label{fig:DifferenceSpectrum2D3D}
 \end{figure}
%======================================================================
%
In \Subsect{SuperfluidTurbulence}, we have defined occupation numbers corresponding to the incompressible, compressible and quantum pressure components of the kinetic energy. 
Numerically, the resulting spectra $n_\mathrm{i}(k)$, $n_\mathrm{c}(k)$, and $n_\mathrm{q}(k)$ add up to $n_{\mathrm{s}}(k)$ which is equal to the full single-particle spectrum $n(k)$ up to small corrections. 
In particular, the scaling behavior emerges as the same for both, $n_{\mathrm{s}}(k)$ and $n(k)$.
To make this more explicit, we calculate $k^2 n(k)$ in terms of generalized velocities $\vector{w}_{\delta}, \delta\in\{$v,q$\}$,
\begin{align}
k^2 n(k) 
=&  \langle \, \mathcal{F}(\vector{w}_{\mathrm{v}} e^{i\varphi})(k) \, \mathcal{F}(\vector{w}_{\mathrm{v}}e^{i\varphi})^*(k) \, \rangle  
\nonumber \\
&+  2\Re \langle  \,  i \mathcal{F}(\vector{w}_{\mathrm{v}}e^{i\varphi})(k) \, \mathcal{F}(\vector{w}_{\mathrm{q}}e^{i\varphi})^*(k) \, \rangle  \nonumber \\
&+  \langle  \, \mathcal{F}(\vector{w}_{\mathrm{q}}e^{i\varphi})(k) \,  \mathcal{F}(\vector{w}_{\mathrm{q}} e^{i\varphi})^*(k)  \, \rangle  \,. 
\end{align}
$\Re$ denotes the real part and $\mathcal{F}$ the Fourier transform. 
The middle term vanishes since the expectation value is purely imaginary due to isotropy. 
It follows that
\begin{align}
k^2& n(k)  
\nonumber\\
&= \langle \, (\vector{w}_{\mathrm{v}} * \mathcal{F}(e^{i\varphi}))(k) \, (\vector{w}_{\mathrm{v}} *\mathcal{F}(e^{i\varphi}))^* (k) \, \rangle 
\nonumber \\
&+ \langle \, (\vector{w}_{\mathrm{q}} * \mathcal{F}(e^{i\varphi}))(k) \, (\vector{w}_{\mathrm{q}} *\mathcal{F}(e^{i\varphi}))^* (k) \, \rangle \,. 
\end{align}
Since phase and phase velocity are expected not to be correlated, the main contribution to the 4-point correlation functions is assumed to arise from terms of the form $ \langle \vector{w}_{\delta}(|\mathbf{p}-\mathbf{k}|) \vector{w}_{\delta}(|\mathbf{q}-\mathbf{k}|) \rangle \langle \mathcal{F}(\exp\{i\varphi\})(p)\mathcal{F}(\exp\{i\varphi\})^*(q) \rangle$, $\delta\in\{$v,q$\}$. In the superfluid regime, the phase of the field is a slowly varying function in position space. Therefore, $\mathcal{F}(\exp\{i\varphi\})^*(q)$-terms are strongly peaked at zero momentum, effectively acting as  regularized delta functions under the convolution. Hence,
\begin{eqnarray}
 k^2 n(k) \simeq  \langle |\vector{w}_{\mathrm{v}}(k)|^2 \rangle + \langle |\vector{w}_{\mathrm{q}}(k) |^2 \rangle.
  \label{eq:approx}
\end{eqnarray}
This approximative identity is supported by \Fig{DifferenceSpectrum2D3D}.
%
%
% Create the reference section using BibTeX:
%\bibliography{Bibliography,additions,mybib}

\end{document}